\documentclass[useAMS,usenatbib]{mn2e}

\usepackage{natbib}
\usepackage{ graphicx}
\usepackage{ subfigure}
\usepackage{multirow}

\title[Black hole-host scaling relations]{Scaling relations between black holes and their host galaxies: comparing theoretical and observational measurements, and the impact of selection effects}
\author[C. DeGraf et al.]  {C. DeGraf$^{[1,2]}$ T. Di Matteo$^{[2]}$, T. Treu$^{[3,4]}$, Y. Feng$^{[2]}$, J.-H. Woo$^{[5]}$, D. Park$^{[6]}$ \\
{1} {Center for Astrophysics and Planetary Science, Racah Institute of Physics, The Hebrew University, Jerusalem 91904, Israel} \\
{2} {McWilliams Center for
       Cosmology, Carnegie Mellon University, 5000 Forbes Avenue, Pittsburgh,
       PA 15213, USA} \\
{3} {Department of Physics, University of California, Santa Barbara, CA 93106, USA} \\
{4} {Department of Physics and Astronomy, University of California, Los Angeles, CA 90095} \\
{5} {Astronomy Program, Department of Physics and Astronomy, Seoul National University, Seoul, 151-742, Republic of Korea} \\
{6} {National Astronomical Observatories, Chinese Academy of Sciences, Beijing 100012, China; EACOA fellow}} 
\def\simgt{\lower.5ex\hbox{$\; \buildrel > \over \sim \;$}}

\begin{document}

\date{Submitted to MNRAS}
\pagerange{\pageref{firstpage}--\pageref{lastpage}}
\pubyear{20??}

\maketitle
\begin{abstract}
We use the high-resolution simulation \textit{MassiveBlackII} to examine scaling relations 
between supermassive black hole mass ($M_{\rm{BH}}$) and their host galaxies' properties 
($\sigma$, total $M_*$ and $L_{\rm V}$), finding good agreement with recent observational data, 
especially at the high-mass end.  The simulations have less intrinsic scatter than observations, 
and the $M_{\rm{BH}}-L_V$ correlation has the largest scatter, suggesting it may the the least 
fundamental of the three relations.We find Gaussian scatter about all three relations, except 
among the highest mass galaxies, which host more massive black holes. Below $z \sim 2$ the slopes 
for the full population remain roughly $z$-independent, and only steepen by $50\%$ by $z \sim 4$. 
The normalization of the $\sigma$, $L_V$ relations evolve by 0.3, 0.43 dex, while the $M_{\rm{BH}}$ 
correlation does not evolve out to at least $z \sim 2$.  Testing for selection biases, we find 
samples selected by $M_{\rm{BH}}$ or $M_*$ have steeper slopes than randomly selected samples. 
If unaccounted for, such a selection function would find faster evolution than inferred 
from a randomly selected sample, as objects at the high-end of the relation tend to evolve 
more rapidly. We find a potential bias among high-$L_{\rm{BH}}$ subsamples (tending to reside 
in higher mass galaxies), but these bright-AGN exhibit no intrinsic bias relative to fainter 
ones in equivalent-mass hosts, nor is there a significant difference between active- and 
inactive- samples.  Finally we characterize the evolution of individual black holes along the 
scaling planes.  Below the local relation, black holes grow faster than their host (72\% of 
black holes $>0.3$ dex below the mean relation have a $M_{\rm{BH}}-M_*$ trajectory steeper than 
the local relation), while those above have shallower trajectories (only 14\% are steeper than 
local).  Thus black holes tend to grow faster than their hosts until surpassing the local 
relation, at which point their growth is suppressed while their hosts continue to grow, returning 
them to the mean relation.

\end{abstract}


\section{Introduction}
\label{sec:Introduction}
In the local universe, the discovery of close relationships between
the masses of supermassive black holes and several properties of their
bulges such as the stellar mass \citep{Magorrian1998, HaringRix2004},
stellar velocity dispersion (${M_{\rm BH}}-\sigma$ relation,
\citet{Gebhardt2000, Ferrarese2002, Tremaine2002, Ferrarese2002,
  Gultekin2009, McConnell2011, McConnellMa2012}, and the concentration
parameter \citep{Graham2001} have revolutionized our view of BHs,
linking their growth to that of its host galaxy \citep[for a recent
review see also][]{KormendyHo2013}.  Most recently,
\citet{McConnellMa2012} have re-analyzed the scaling relations and
added a number of high mass objects.  To understand the evolution of
these relations at higher redshifts (mostly up to $z \sim 2$)
observational studies rely on galaxies with active galactic nuclei
\citep{Merloni2010, Bennert2010, Bennert2011, Bennert2011b,
  KormendyHo2013, Park2014} 
for which BH mass
estimates use the so-called virial method \citep{Wandel1999}.  
Whereas some studies have found an evolution in which BH growth precedes that of their host galaxies, others found no evolution within the uncertainties.  These observations are challenging and their interpretation relies on a full understanding of systematic uncertainties \citep[in both the BH mass and host galaxy measurements, e.g. ][]{Woo2006} and selection effects \citep{Treu2007, Lauer2007, SchulzeWisotzki2011, SchulzeWisotzki2014}.  Without a proper modeling of the uncertainties, selection effects at high redshift can lead to much stronger apparent evolution than the true population undergoes, and underestimate the slope of the relations \citep{Volonteri2011}.

A popular way to interpret these
relationships is by assuming that supermassive BHs regulate their own
growth and that of their hosts by coupling some (small) fraction of
their energy output to their surrounding gas. This, so called, ``AGN
feedback'' and heats and unbinds significant fractions of
the gas and inhibits star formation \citep{SilkRees1998, Springel2005,
  Bower2006, Croton2006, DiMatteo2008, Ciotti2009, Fanidakis2011}. The
scaling relations of black hole mass with the stellar properties of
the host galaxies have important implications for black hole and galaxies as well as
understanding the importance and the effects of AGN feedback.

Here we use state-of-the-art cosmological hydrodynamical simulations
of structure formation ({\em MassiveBlack-II}) \citep{Khandai2014} to
investigate the predictions of the galaxy-black hole relations
$M_{\rm{BH}}-\sigma$, $M_{\rm BH}-M_{*, tot}$ and $M_{\rm BH}-L_{V,
  tot}$ relations for the population of black holes and compare them
to the observational constraints at $z=0-2$.  MBII is a recent large-scale and high resolution hydrodynamic
simulation in a box of $100 \rm{Mpc}/h$ on the side, making one of the
largest cosmological Smooth Particle Hydrodynamics (SPH) simulation to
date with “full physics” of galaxy formation (meaning here an
inclusion of radiative cooling, star formation, black hole growth and
associated feedback physics). Note that \citet{Sijacki2014} has
also presented results on the black hole scaling relations from the
Illustris moving-mesh simulation \citep{Vogelsberger2014}, also well matching the local relation and finding significant redshift evolution.

We
concentrate on the prediction for the black hole-galaxy relations
relative to the the {\it total} stellar mass and luminosity of the
host galaxy as measured by \citet{Merloni2010, Bennert2010,
  Bennert2011, Bennert2011b, KormendyHo2013, Park2014}.  The
predictions from the simulations are most direct for the total
quantities (stellar mass and luminosity) and we do not need to
indroduce unknown biases as we would have if we used proxies for these
measurements in galaxy bulge components (we reserve this work to a
future analysis with morphological decomposion). The aim is then to
first clearly understand these different relations (e.g. which one may
be the strongest correlation), as well as predict the expected
redshift evolution for $M_{\rm BH}-M_{*, tot}$ and $M_{\rm BH}-L_{V,
  tot}$ planes. 
We will also address the issue of selection effects in observing these relations.  Since the observational investigations into
scaling relations at $z > 0$ use AGN-selected samples with luminosity
limitations, understanding the effect $L_{\rm{BH}}$ and $M_{\rm{BH}}$ has on the scaling
relations is also significant.

The paper is organized as follows: In Section \ref{sec:Method} we will
briefly describe MBII simulation and the black hole model contained
therein.  In Section \ref{sec:Results} we investigate the 3 primary
scaling relations and their redshift evolution (\S \ref{sec:msigma}
and \ref{sec:mmstar_mLV}), the dependence on black hole luminosity (\S
\ref{sec:L_dependence}) and the evolution of typical black hole mass
(\S \ref{sec:offset_evolution}). In Section \ref{sec:subsampling} we
discuss how the distribution of objects in smaller subsamples selected
by $M_*$ or $M_{\rm BH}$ can bias the inferred slope and evolution of the
scaling relations, and in Section \ref{sec:tracing_evolution} we probe
the typical evolution of individual black holes on the scaling
relation planes.  We summarize our results in Section
\ref{sec:Conclusions}.

\section{Method}
\label{sec:Method}

In this paper we use a cosmological hydrodynamic simulation
\textit{MassiveBlackII} \citep{Khandai2014}. This simulation is
similar (in terms of the physics modelled) to the high-redshift $533
\: h^{-1}$ Mpc \textit{MassiveBlack} simulation, using a smaller (100
$h^{-1}$ Mpc) box, but allowing for higher-resolution and a complete
run to $z \sim 0$ (see Table \ref{simparam}). These simulations have
been performed with the cosmological TreePM-Smooth Particle
Hydrodynamics (SPH) code {\small P-GADGET}, a {\it hybrid} version of
the parallel code {\small GADGET2} \citep{Springel2005d} which has
been extensively modified and upgraded to run on the new generation of
Petaflop scale supercomputers. The major improvement over previous
versions of {\small GADGET} is in the use of threads in both the
gravity and SPH part of the code which allows the effective use of
multi core processors combined with an optimum number of MPI task per
node.  The {\it MassiveBlackII} simulation contains $N_{\it part} = 2
\times 1792^3 = 11.5$ billion particles in a volume of $100 \rm{Mpc}
/h$ on a side with a gravitational smoothing length $\epsilon = 2.0
\rm{kpc}/h$ (in comoving units). The gas and dark matter particle
masses are $m_{\rm g} = 2.1 \times 10^6 M_\odot $ and $m_{\rm DM} = 1
\times 10^7 M_\odot$ respectively.  The simulation has currently been
run from $z=159$ to $z=0.06$.  

The run contains the standard gravity and hydrodynamics, as well as
additional (subgrid) modeling for star formation
\citep{SpringelHernquist2003}, black holes and associated feedback
processes \citep{DiMatteo2008, DiMatteo2012}. The cosmological
parameters used were: the amplitude of mass fluctuations,
$\sigma_8=0.8$, spectral index, $n_s = 0.96$, cosmological constant
parameter $\Omega_{\Lambda}= 0.725$, mass density parameter $\Omega_m
= 0.275$ , baryon density parameter $\Omega_b = 0.044$ and $h=0.702$
(Hubble's constant in units of $100 \mathrm{km\:s}^{-1}
\mathrm{Mpc}^{-1}$), based on WMAP7.

Within our simulation, black holes are modeled as collisionless sink
particles which form in newly emerging and resolved dark matter halos.
These halos are found by calling a friends of friends group finder at
regular intervals (in time intervals spaced by $\Delta \log a = \log
1.25$).  Any group above a threshold mass of $5 \times 10^{10} h^{-1}
M_\odot$ not already containing a black hole is provided one by
converting its densest particle to a sink particle with a seed mass of
$M_{\rm{BH,seed}} = 5 \times 10^5 h^{-1} M_\odot$.  This seeding
prescription is chosen to reasonably match the expected formation of
supermassive black holes by gas directly collapsing to BHs with
$M_{\rm{BH}} \sim M_{\rm{seed}}$ \citep[e.g.][]{BrommLoeb2003,
  Begelman2006} or by PopIII stars collapsing to $\sim 10^2 M_\odot$
BHs at $z \sim 30$ \citep{Bromm2004, Yoshida2006} followed by
sufficient exponential growth to reach $M_{\rm{seed}}$ by the time the
host halo reaches $\sim 10^{10} M_\odot$.  Following insertion, BHs
grow in mass by accretion of surrounding gas and by merging with other
black holes.  Gas is accreted according to $\dot{M}_{\rm BH} = \alpha
\frac {4 \pi G^2 M_{\rm BH}^2 \rho}{(c_s^2 + v^2)^{3/2}}$
\citep{1939PCPS...35..405H, 1944MNRAS.104..273B, 1952MNRAS.112..195B},
where $\rho$ is the local gas density, $c_s$ is the local sound speed,
$v$ is the velocity of the BH relative to the surrounding gas, and
$\alpha$ is introduced to correct for the reduction of the gas density
close to the BH due to our effective sub-resolution model for the
ISM. To allow for the initial rapid BH growth necessary to produce
sufficiently massive BHs at early time ($\sim 10^9 M_\odot$ by $z \sim
6$) we allow for mildly super-Eddington accretion \citep[consistent
with][]{VolonteriRees2006, Begelman2006}, but limit it to a maximum of
$3 \times \dot{M}_{\rm{Edd}}$ to prevent artificially high values.

\begin{table}
\caption{Numerical Parameters for the \textit{MassiveBlackII} simulation}
\begin{tabular}{c c c c c}
  \hline
  \hline
Boxsize & $N_p$ & $m_{DM}$ & $m_{gas}$ & $\epsilon$ \\
 $h^{-1}$ Mpc & & $h^{-1} M_\odot$ &  $h^{-1} M_\odot$ &  $h^{-1}$ kpc \\

\hline

100 & $2 \times 1792^3$ & $1 \times 10^7$ & $2.1 \times 10^6$ & 2.0 \\

\hline

\end{tabular}
\label{simparam}
\end{table}

The BH is assumed to radiate with a bolometric luminosity proportional to the
accretion rate, $L = \eta \dot{M}_{\rm{BH}} c^2$ \citep{ShakuraSunyaev1973},
where the radiative efficiency $\eta$ is fixed to 0.1 throughout the
simulation and our analysis. To model the expected coupling between the
liberated radiation and the surrounding gas, 5 per cent of the luminosity is
isotropically deposited to the local black hole kernel as thermal energy.  The
5 per cent value for the coupling factor is based on galaxy merger simulations
such that the normalization of the $M_{\rm{BH}}-\sigma$ relation is reproduced
\citep{DiMatteo2005}.

The second mode of black hole growth is through mergers which occur when dark
matter halos merge into a single halo, such that their black holes fall toward
the center of the new halo, eventually merging with one another.  In
cosmological volumes, it is not possible to directly model the physics of the
infalling BHs at the smallest scales, so a sub-resolution model is used.
Since the mergers typically occur at the center of a galaxy (i.e. a gas-rich
environment), we assume the final coalescence will be rapid
\citep{MakinoFunato2004, Escala2004, Mayer2007}, so we merge the BHs once they
are within the spatial resolution of the simulation.  However, to prevent
merging of BHs which are rapidly passing one another, mergers are prevented if
the BHs' velocity relative to one another is too high (comparable to the local
sound speed).

The model used for black hole creation, accretion and feedback has been
investigated and discussed in \citet{Sijacki2007, DiMatteo2008, Colberg2008,
Croft2009, Sijacki2009, DeGraf2010, DeGrafClustering2010}, finding it does a
good job reproducing the $M_{\rm{BH}}-\sigma$ relation, the total black hole
mass density \citep{DiMatteo2008}, the QLF \citep{DeGraf2010}, and the
expected black hole clustering behavior \citep{DeGrafClustering2010}. This
simple model thus appears to model the growth, activity, and evolution of
supermassive black holes in a cosmological context surprisingly well (though
the detailed treatement of the accretion physics is infeasible for
cosmological scale simulations).  We also note that \citet{BoothSchaye2009}
and \citet{Johansson2008} have adopted a very similar model, and have
independently investigated the parameter space of the reference model of
\citet{DiMatteo2008}, as well as varying some of the underlying prescriptions.
For further details on the simulation methods and convergence studies done for
similar simulations, see \citet{DiMatteo2008}.

\begin{figure}
\centering
\includegraphics[width=8cm]{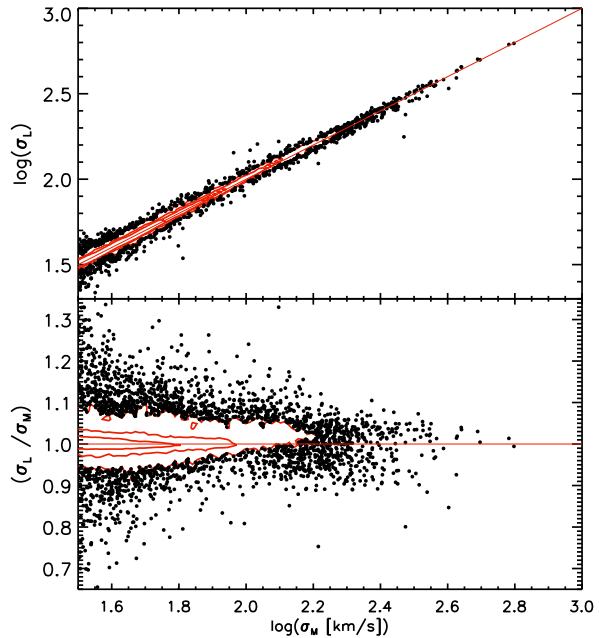}
\caption{Direct comparison between velocity dispersion within
  half-mass radius ($\sigma_M$) and within half-light radius
  ($\sigma_L$).  Top panel shows direct comparison between the two
  dispersions; bottom panel shows the ratio $\sigma_L/\sigma_M$ as a
  function of $\sigma_M$.}
\label{sigmacompare}
\end{figure}

Because the simulation saves the complete set of black hole properties (mass,
accretion rate, position, local gas density, sound speed, velocity, and BH
velocity relative to local gas) for each BH at every timestep, the black hole
output for such a large simulation is prohibatively difficult to analyze using
previous techniques.  For this reason, \citet{Lopez2011} developed a
relational database management system specifically for this simulation.  A
similar strategy has also been followed in the analysis of the Millenium
simulation \citep{Lemson2006}. In addition to providing a substantially more
efficient query system for extracting information, this database is
significantly more flexible than traditional approaches.  For a complete
summary of the database format and its efficiency, please see
\citet{Lopez2011}.

Catalogues of galaxies are made from the simulation outputs by first using
a friends-of-friends groupfinder and then applying the {\sc SUBFIND} algorithm
(Springel 2001) to find gravitationally bound subhalos. The stellar
component of each subhalo consists of a number of star particles, each labelled
with a mass and the redshift at which the star particle was created.

For our galaxies the spectral energy distribution (SED) of a galaxy is
generated by summing the SEDs of each star particle in the galaxy.
The SED of the star particles is generated using the Pegase.2 stellar
population synthesis (SPS) code \citep{Fioc1997, Fioc1999} by
considering their ages, mass and metallicities and assuming a Salpeter
IMF.  Nebula (continuum and line) emission is also added to each
star. With stellar luminosity we have information about direct
half-light radii of galaxies and can carry out a luminosity-weighted
velocity dispersion for our subgroups catalog. To most closely match
observational studies, for this work we use the V-band rest frame as
our luminosity band.

\begin{figure*}
\centering
\includegraphics[width=15cm]{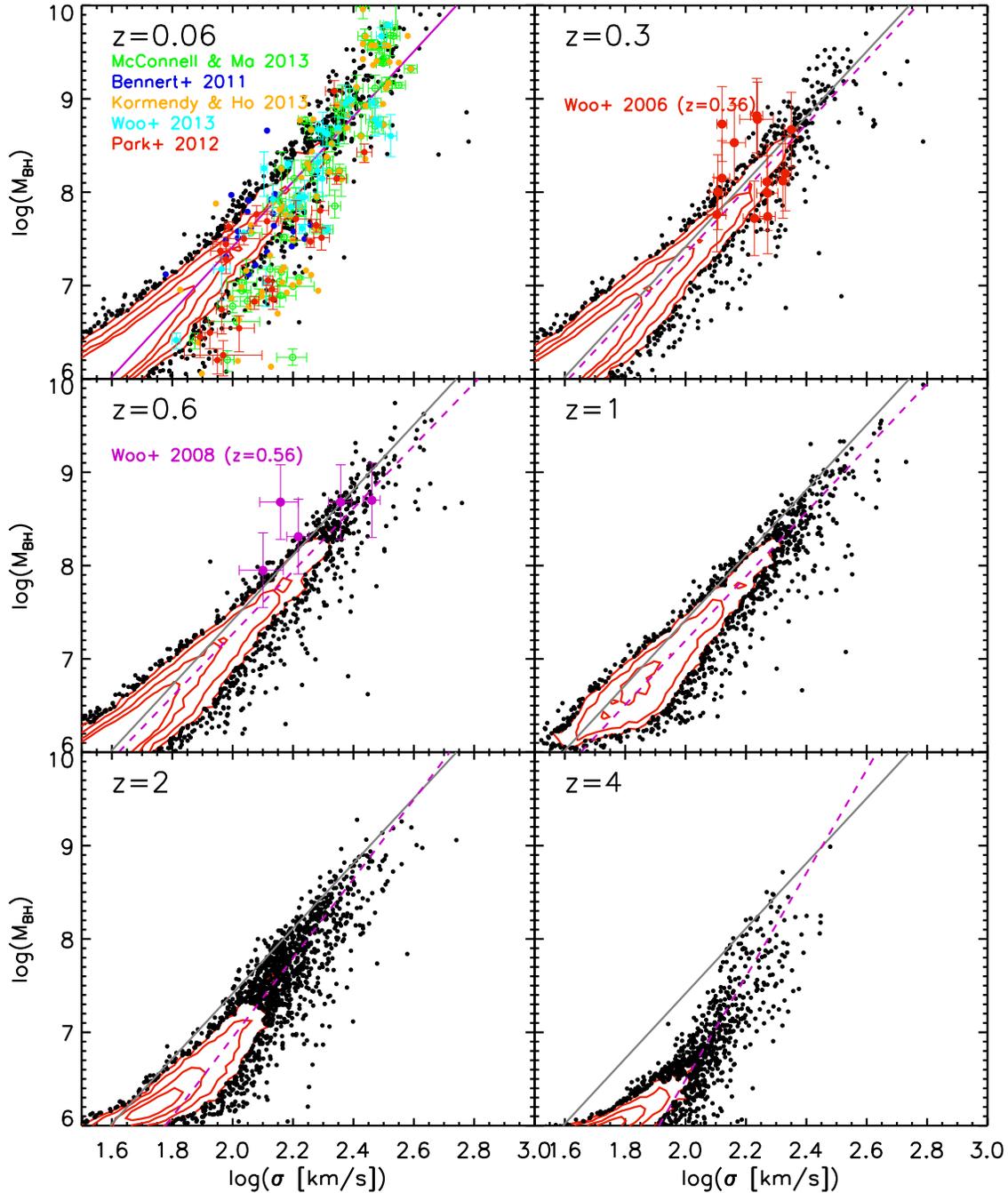}
\caption{Black hole mass vs. luminosity-weighted stellar velocity dispersion within the half-light radius (black datapoints and red contours).  The grey line shows the best-fitting relation calculated at $z=0.06$, and is included at all redshifts for reference. Observational data are provided as colored datapoints.  Note: for the \citet{Woo2006,Woo2008} data, we use the updated BH mass estimates from \citet{Park2014}.}
\label{fig:mbh_sigma}
\end{figure*}

\begin{figure*}
\centering
\includegraphics[width=15cm]{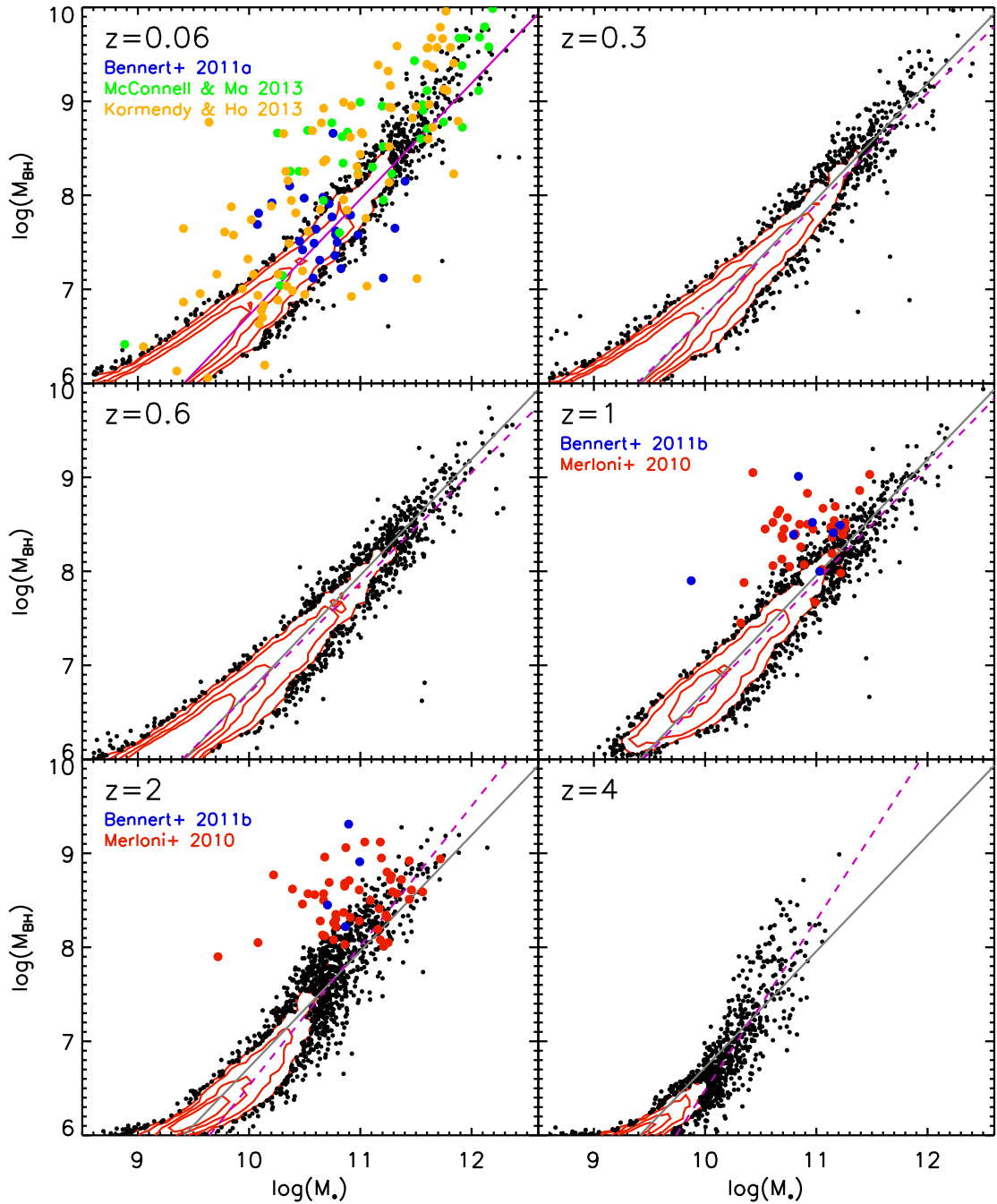}
\caption{Black hole mass vs. stellar mass within the 2 times the V-band half-light radius (black datapoints and red contours) compared to observational data (colored datapoints).  The grey line shows the best-fitting relation calculated at $z=0.06$, and is included at all redshifts for reference.  Note that the data from \citet{McConnellMa2012} and \citet{KormendyHo2013} are bulge mass, not total mass, and should thus be considered lower-limits for the $M_*$ comparison.}
\label{fig:mbh_mstar}
\end{figure*}

\begin{figure*}
\centering
\includegraphics[width=15cm]{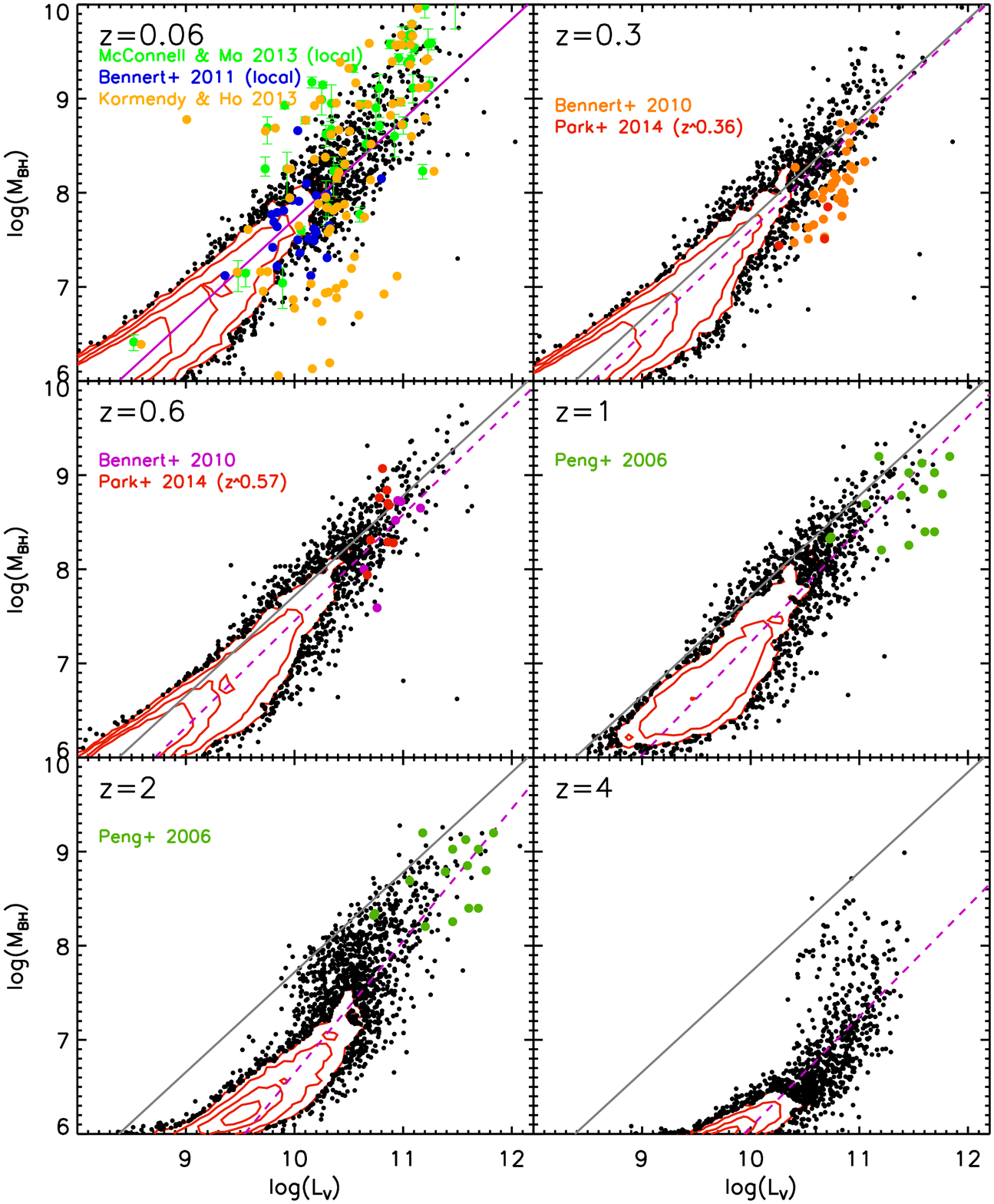}
\caption{Black hole mass vs. observed V-band luminosity within 2 times the half-light radius (black datapoints and red contours) compared to observational data (colored datapoints).  The grey line shows the best-fitting relation calculated at $z=0.06$, and is included at all redshifts for reference.}
\label{fig:mbh_LV}
\end{figure*}

\section{The scaling relations}
\label{sec:Results}

\subsection{$M_{\rm BH} - \sigma$}
\label{sec:msigma}

Figure \ref{sigmacompare} compares $\sigma_L$, the velocity dispersion
within the half-light radius, to $\sigma_M$, within the half-mass
radius (a proxy commonly adopted in simulation for the bulge velocity
dispersion).  Although there is noticeable scatter between the two
calculations, there is no systematic offset between $\sigma_L$ and
$\sigma_M$ (even at lower-$\sigma$ the scatter remains $\sim 20\%$).
This lack of a systematic offset suggests that either proxy is adequate given the scatter of the relations; however we caution that possible residual systematics at the 20\% level might be present (as mass
and light trace each other at these scales).  Furthermore, this proxy neglects the addition of rotational velocity, and is more uncertain at the low mass end \citet[see ][]{Sijacki2014}.  This precise impact is investigated in detail in an upcoming paper on bulge-decomposition in the simulation \citep{Tenneti2014}.  For the rest of this
analysis we choose to work with $\sigma_L$, since the
luminosity-radius is more similar to observational approaches.  Note
that for the remainder of this paper we refer only to $\sigma$ (with
no subscript), but in all cases it is used to refer to the V-band
luminosity-weighted velocity dispersion within the V-band half-light
radius.

In Figure \ref{fig:mbh_sigma} we show the $M_{\rm{BH}}-\sigma$
relation from our simulation in the redshift range 0.06-4.  These results are
shown as black datapoints, with red contours representing the regions
of highest concentration (to show the behavior where the concentration
of points is too high).  Note that we only consider central black
holes \citep[see also][]{Sijacki2014}.  Observational data from
\citet{Woo2006, Woo2008, Bennert2011, McConnellMa2012, KormendyHo2013}
are shown as colored points.  We also provide a best-fitting relation
for our black holes (dashed pink line).  For reference, the
best-fitting relation from the $z$=0.06 panel is shown in all panels as
a solid grey line.  The best-fitting lines are calculated using the
functional form
\begin{equation}
\frac{M_{BH}}{M_\odot}=10^\alpha \left ({\frac{x}{x_0}} \right )^\beta
\label{eqn:fit}
\end{equation}
(where $x=\sigma, M_*, L_V$, etc.).  Following
\citet{McConnellMa2012}, this fitting is accomplished using the
least-squares fitting routine MPFITEXY \citep[see
][]{Williams2010}. Each black hole has an uncertainty in $\sigma$
($\epsilon_\sigma$, set to the standard deviation of the logarithm of
the projected dispersion for the three orthogonal directions).
$M_{\rm{BH}}$, $M_*$, and $L_V$ do not have equivalent variations to
use, so a 5\% uncertainty is assigned to each (though the fits are not
sensitive to this value).  To avoid the fit being dominated by the
low-mass objects (which are less well-resolved and more dependent on
the BH seeding prescription), we limit our fits to black holes within
hosts with $M_* > 10^{10} M_\odot$ (where $M_*$ is the total stellar mass).
The parameters $\alpha$ and $\beta$ for these
fits are provided in Table \ref{table:mainparams}.  Following
\citet{Gultekin2009} and \citet{McConnellMa2012} we include a measure
of the intrinsic scatter ($\epsilon_0$), such that
\begin{equation}
\chi^2 = \sum_i \frac{\left ( \rm{log}_{10} \left ( M \right ) - \alpha - \beta
    \rm{log}_{10} \left ( \sigma_i / 200 \rm{km/s}
\right ) \right )^2}{\epsilon_0^2 + \epsilon_{M_{\rm{BH},i}}^2 +
\beta^2 \epsilon_{\sigma,i}^2}
\end{equation}
equals the number of degrees of freedom.  This intrinsic scatter
presents an important component of any scaling analysis, and must be
considered when performing the fitting.

When comparing with observational data, we find that our simulation
does an excellent job matching the observational data at the high-end
(above $\sim 10^{7.3} M_\odot$), including the scatter within the
relation.  At lower-masses and correspondingly low $\sigma$ (at least
at $z=0$, where such low-mass objects have been observed), our
simulations either somewhat overestimate the mass of the black holes
in a given galaxy or underestimate $\sigma$.  Note also that
\citet{Sijacki2014} has shown that the effect of adding the rotational
velocity component (as typically included in observations) to $\sigma$
has a strong effect at this end of the relation. In particular, they
found that the hosts of low-mass black holes can have a significant
rotational component, leading to a larger $\sigma$ than when using a
pure velocity dispersion and bringing our results closer to the
observed data.  This effect will be investigated in an upcoming work
(involving a complete bulge decomposition), but given that we will
concentrate on the relations with the total $M_*$ and $L_V$ rather
than bulge proxies (see Section \ref{sec:mmstar_mLV}) in this
analysis, we will not further need to consider these effects here.

Of particular interest from our simulation is the redshift evolution
of the relation, as we have far larger samples and dynamic range at
high-$z$ than can be expected from any current observational study.
Of particular note in our simulation is that the $M_{\rm{BH}}-\sigma$
relation has minimal evolution out to $z \sim 1$, with the slope
changing by less than 10\% \citep[similar to the findings of the Magneticum Pathfinder Simulation, ][]{Hirschmann2014, Bachmann2014}, and the normalization by only 0.2 dex (see
also Section~\ref{sec:subsampling}). At $z=2-4$ we find somewhat more
evolution, with the typical black hole mass being smaller, and the
slope being steeper.  This is consistent with the `selective accretion' prediction of \citet{Volonteri2011}, that accretion is more efficient in high-mass halos at high redshift, while at low redshift baryonic processes wash out this dependence on halo mass. However, we note that this is due, at least in
part, to the more-recently seeded black holes.  Particularly at $z=4$,
we see that in the low-$\sigma$ hosts (below $\sigma \sim 100$ km/s),
the relation begins to flatten close to $10^6 M_\odot$, where we are
close to our seed mass.

\subsection{$M_{\rm BH} - M_*$ and $M_{\rm BH} - L_V$}
\label{sec:mmstar_mLV}
In Figures \ref{fig:mbh_mstar} and \ref{fig:mbh_LV} we show the
$M_{\rm{BH}}-M_*$ and $M_{\rm{BH}}-L_V$ relations, respectively.  We
calculate the stellar mass ($M_*$) and V-band luminosity ($L_V$)
within 2 times the V-band half-light radius (this definition does not
differ from the total in low/intermediate mass systems but it allows
us to adequately exclude some of the intracluster light for the
massive systems, \citet[see also][]{Vogelsberger2014}).  As in Figure
\ref{fig:mbh_sigma}, we provide the best fitting relation as a dashed
pink line (with the $z=0.06$ fit shown in all panels in grey for
reference), and observational measurements as colored datapoints.
When considering current observational measurements for the total
relation the agreement is very good with the simulations at $z=0$.
However the scatter in observations is large whilst simulations
predict a rather tight relation, with our intrinsic scatter approximately half that of \citet{McConnellMa2012} \citep[note other simulations have also found a tigher correlation than observations, e.g. ][]{RagoneFigueroa2013}.  We find a well-defined relation with
small scatter ($\epsilon_0 \sim 0.18$), minimal evolution out to
redshift 1, and at $z=2$,$4$ the relation is found to be progressively
steeper. We note however that \citet{Bennert2011, Merloni2010} appear
to find a large population of high-$M_{\rm{BH}}$, moderate-$M_*$
objects which are offset from the simulation relation, particularly at
$z=1$ and 2. 
The disagreement at these redshifts could be due to an unaccounted for selection bias in the observations (biasing the results toward higher-mass black holes), systematic uncertainties in the virial black hole measurements at these redshifts, or a problem with the simulation underproducing sufficiently large black holes at $z \sim 1$ and above.
This is somewhat hard to
envisage as it would require a some population of black holes which are
capable of growing almost 2 dex above the local relation without reaching a self-regulating phase that slows the BH growth (in sharp contrast to the evolution we find in Section \ref{sec:tracing_evolution}).  We note
that at $z=0.06$, the data from \citet{McConnellMa2012} and
\citet{KormendyHo2013} both have a population that lies above our
black holes, contributing to the much larger scatter in the observations than in the simulations.  This
high-mass population is due, at least in part, to the fact that both
\citet{McConnellMa2012} and \citet{KormendyHo2013} use the stellar
mass of the bulge (for both elliptical and spiral galaxies), rather than the full stellar mass we use.  In particular, we note that the low-mass objects from \citet{KormendyHo2013} where we have the strongest discrepancy are the spiral and S0 type galaxies (where the bulge mass is a significant underestimate), while the high-mass objects are ellipticals, suggesting that correcting for the total mass may help solve this discrepancy.  This will be investigated in more detail in an upcoming work involving a complete bulge-decomposition \citep{Tenneti2014}.  The
second difference from Figure \ref{fig:mbh_sigma} is that the evolution
of the relation to $z=2$,4 is notably different than that of
$M_{\rm{BH}}-\sigma$.  Unlike the $M_{\rm{BH}}-\sigma$ relation, we
find that the high-mass objects at $z=2$,4 tend to be \textit{larger}
with respect to their host mass than a comparable object at
lower-redshift, telling us that at high-redshift, the black holes are
being fuelled more efficiently (compared to their host galaxy) than a
similar object at low-redshift.  This is consistent with previous
high-redshift growth rates found in these simulations
\citep{DeGrafBHgrowth2012}.

Finally, Figure~\ref{fig:mbh_LV} shows the BH mass relative to the
host luminosity. Here we note several important differences with
respect to the previous relations.  Firstly, the $M_{\rm{BH}}-L_V$
relation has an intrinsic scatter approximately $50\%$ larger than
either $M_{\rm{BH}}-\sigma$ or $M_{\rm{BH}}-M_*$, consistent at all
redshifts, suggesting that the relation with $\sigma$ and with $M_*$
are more fundamental than $L_V$. This is consistent with the findings
of \citet{McConnellMa2012, KormendyHo2013}.  Secondly, we find significantly
more evolution in the typical black hole mass for a given host
luminosity. 
Rather than an inherent change in the BH-host relation, this can be interpreted as an evolution in the typical mass-to-light ratio of the host galaxy, since high-$z$ galaxies tend to host younger, and therefore brighter, stellar populations.
This effect is discussed in more detail (including using an evolution-corrected luminosity $L_{V,0}$) in Section \ref{sec:offset_evolution} while investigating evolution of the typical black hole mass.

\begin{table}
\centering
\begin{tabular}{c c c c}
& && \\
\multicolumn{4}{c}{Parameters for scaling relation} \\

\hline
\hline
 $z$  & $\alpha$ & $\beta$ & $\epsilon_0$ \\
\hline
\multicolumn{4}{c}{$M_{\rm{BH}}-\sigma$ relation ($x_0 = 200 \rm{km/s}$)}\\
\hline
 0.06 & $8.465 \pm 0.007$ & $3.487 \pm 0.019$ & $0.169$  \\
 0.3 & $8.317 \pm 0.006$ & $3.295 \pm 0.019$ & $0.163$  \\
 0.6 & $8.206 \pm 0.007$ & $3.186 \pm 0.020$ & $0.179$  \\
 1 & $8.154 \pm 0.007$ & $3.248 \pm 0.022$ & $0.192$  \\
 2 & $8.091 \pm 0.011$ & $3.852 \pm 0.037$ & $0.235$  \\
 4 & $7.860 \pm 0.030$ & $4.416 \pm 0.137$ & $0.292$  \\

\hline
\multicolumn{4}{c}{$M_{\rm{BH}}-M_*$ relation ($x_0 = 10^{11} M_\odot$)}\\
\hline
 0.06 & $7.957 \pm 0.004$ & $1.235 \pm 0.007$ & $0.184$  \\
 0.3 & $7.903 \pm 0.004$ & $1.183 \pm 0.006$ & $0.175$  \\
 0.6 & $7.878 \pm 0.004$ & $1.174 \pm 0.007$ & $0.179$  \\
 1 & $7.891 \pm 0.005$ & $1.213 \pm 0.008$ & $0.189$  \\
 2 & $8.014 \pm 0.009$ & $1.486 \pm 0.013$ & $0.231$  \\
 4 & $8.290 \pm 0.040$ & $1.815 \pm 0.053$ & $0.291$  \\
\hline
\multicolumn{4}{c}{$M_{\rm{BH}}-L_V$ relation  ($x_0 = 10^{10.5} L_\odot$)}\\
\hline
 0.06 & $8.249 \pm 0.008$ & $1.062 \pm 0.008$ & $0.258$  \\
 0.3 & $8.155 \pm 0.008$ & $1.106 \pm 0.009$ & $0.256$  \\
 0.6 & $8.016 \pm 0.008$ & $1.129 \pm 0.010$ & $0.264$  \\
 1 & $7.823 \pm 0.008$ & $1.198 \pm 0.013$ & $0.285$  \\
 2 & $7.347 \pm 0.008$ & $1.404 \pm 0.022$ & $0.352$  \\
 4 & $6.659 \pm 0.031$ & $1.174 \pm 0.086$ & $0.450$  \\

\end{tabular}
\caption{Best fitting parameters for the scaling relations (Equation
  \ref{eqn:fit}).} 
\label{table:mainparams}
\end{table}

\begin{table}
\centering
\begin{tabular}{c c c c}

& && \\

\multicolumn{4}{c}{Dependency on $L_{\rm{BH}}$, $\sigma$, $M_*$} \\

\hline
\hline
&   $\alpha$ & $\beta$ & $\epsilon_0$ \\
\hline
\hline
\multicolumn{4}{c}{$M_{\rm{BH}}-\sigma$ relation ($x_0 = 200 \rm{km/s}$)}\\
\hline
\hline
$L_{\rm{BH}} > 10^8 L_\odot$ & $8.467 \pm 0.007$ & $3.447 \pm 0.019$ & $0.158$  \\
$L_{\rm{BH}} > 10^9 L_\odot$ & $8.518 \pm 0.009$ & $3.520 \pm 0.034$ & $0.188$  \\
$L_{\rm{BH}} > 10^{10} L_\odot$ & $8.636 \pm 0.013$ & $3.765 \pm 0.079$ & $0.204$  \\
$L_{\rm{BH}} > 10^{11} L_\odot$ & $8.677 \pm 0.033$ & $4.135 \pm 0.218$ & $0.249$  \\
\hline
$\sigma <200 \rm{km/s}$ & $8.681 \pm 0.042$ & $2.688 \pm 0.296$ & $0.351$  \\
$\sigma >200 \rm{km/s}$ & $8.415 \pm 0.008$ & $3.354 \pm 0.022$ & $0.155$  \\
\hline
$M_* <10^{10.5} M_\odot$ & $8.286 \pm 0.026$ & $3.038 \pm 0.060$ & $0.139$  \\
$M_* >10^{10.5} M_\odot$ & $8.511 \pm 0.009$ & $3.827 \pm 0.042$ & $0.199$  \\

\hline
\hline
\multicolumn{4}{c}{$M_{\rm{BH}}-M_*$ relation ($x_0 = 10^{11} M_\odot$)}\\
\hline
\hline
$L_{\rm{BH}} > 10^8 L_\odot$ & $7.965 \pm 0.004$ & $1.222 \pm 0.007$ & $0.176$  \\
$L_{\rm{BH}} > 10^{10} L_\odot$ & $8.093 \pm 0.012$ & $1.298 \pm 0.024$ & $0.203$  \\
\hline
$\sigma <200 \rm{km/s}$ & $8.333 \pm 0.074$ & $0.878 \pm 0.096$ & $0.358$  \\
$\sigma >200 \rm{km/s}$ & $7.931 \pm 0.005$ & $1.196 \pm 0.008$ & $0.171$  \\
\hline
$M_* <10^{10.5} M_\odot$ & $7.824 \pm 0.018$ & $1.054 \pm 0.022$ & $0.160$  \\
$M_* >10^{10.5} M_\odot$ & $7.955 \pm 0.005$ & $1.345 \pm 0.014$ & $0.208$  \\

\hline
\hline
\multicolumn{4}{c}{$M_{\rm{BH}}-L_V$ relation  ($x_0 = 10^{10.5} L_\odot$)}\\
\hline
\hline
$L_{\rm{BH}} > 10^8 L_\odot$ & $8.266 \pm 0.008$ & $1.055 \pm 0.008$ & $0.245$  \\
$L_{\rm{BH}} > 10^{10} L_\odot$ & $8.449 \pm 0.016$ & $1.116 \pm 0.035$ & $0.313$  \\
\hline
$\sigma <200 \rm{km/s}$ & $8.754 \pm 0.046$ & $0.564 \pm 0.092$ & $0.397$  \\
$\sigma >200 \rm{km/s}$ & $8.155 \pm 0.009$ & $0.973 \pm 0.009$ & $0.239$  \\
\hline
$M_* <10^{10.5} M_\odot$ & $7.454 \pm 0.025$ & $0.400 \pm 0.022$ & $0.204$  \\
$M_* >10^{10.5} M_\odot$ & $8.291 \pm 0.010$ & $1.043 \pm 0.016$ & $0.288$  \\
\end{tabular}
\caption{Dependency of the scaling relations (Equation
  \ref{eqn:fit}) on black hole luminosity, host velocity dispersion, and stellar mass.} 
\label{table:paramdep}
\end{table}

\subsection{Redshift evolution of BH-galaxy relations}
\label{sec:offset_evolution}
\begin{figure}
\centering
\includegraphics[width=9cm]{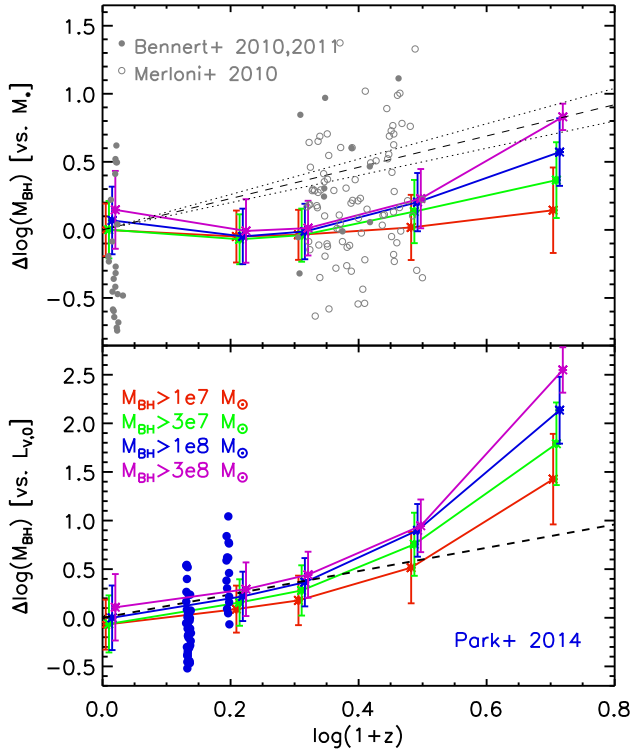}
\caption{\textit{Top Panel:} $M_{\rm{BH}}$ offset with respect to our local $M_{\rm{BH}}-M_*$ relation as a function of redshift for several cuts on minimum black hole mass.  Mass cuts are offset by $\Delta$log(1+$z$)=0.005 to visually separate them.  For comparison, we have observational datapoints (Bennert et al. 2010, 2011, Merloni et al. 2011) and the observed best-fit relation (dashed line, with 1-$\sigma$ range dotted line) from Bennert et al. 2011.  \textit{Bottom Panel:} As in the top panel, but based on the $M_{\rm{BH}}-L_V$ relation, using an evolution-corrected value of $L_{V,0}$ (see text). Grey dashed line is the evolution relation from \citet{Park2014}.}
\label{fig:mass_offset_evolution}
\end{figure}

To provide an accurate estimate for the evolution of a typical black
hole relative to its host, in the top panel of Figure
\ref{fig:mass_offset_evolution} we show the average mass offset
($\Delta(M_{\rm{BH}})$) relative to our best-fitting local
$M_{\rm{BH}}-M_*$ relation (i.e. using the $z=0.06$ best-fitting
parameters from Table \ref{table:mainparams}).  Due to the large
number of black holes in our simulation, we do not show individual
objects, but rather the average offset, with $1-\sigma$ scatter
provided.  We provide four such curves, using four different
lower-limits on black hole mass.  To avoid recently-seeded objects,
our lowest limit is $10^7 M_\odot$, with approximately half-dex
intervals up to $3 \times 10^8 M_\odot$.  Regardless of the mass cut,
we find essentially no evolution to $z \sim 1$.  Above redshift 1, we
begin to see significant mass-dependence, showing that low-mass black
holes don't evolve much, but the high-mass black holes tend to be more
massive relative to their host mass than their low-redshift
counterparts.  This is consistent with Figure \ref{fig:mbh_mstar},
where we saw that the high-mass objects at $z=2$,4 go above the local
relation.  As previously mentioned, this can be explained by the
faster growth found in high-redshift black holes.

We also compare our offset evolution to data from \citet{Bennert2010,
  Bennert2011, Merloni2010}, and the best-fitting relation of
\citet{Bennert2011} (which takes into account the selection function; dashed line).  We note that our evolution is
somewhat less than the best-fitting result of \citet{Bennert2011}, but lies well within the scatter of the data.

In the bottom panel of Figure \ref{fig:mass_offset_evolution} we show
the evolution in mass-offset relative to the $M_{\rm{BH}}-L_V$
relation, rather than $M_{\rm{BH}}-M_*$.  We note that if we were to use the base $L_V$ to calculate $\Delta M_{\rm{BH}}$ we would find drastically
different behavior, with the black holes at higher redshifts tending
to be much less-massive relative to their host luminosities.  
However, this is primarily due to the evolution in host luminosity: high-$z$ galaxies 
host primarily younger stars, and thus tend to be brighter than their low-$z$ counterparts at fixed black hole and stellar mass.
Thus the evolution is dominated by the evolution in the stellar-mass-to-light ratio of the host, rather than interactions with the black hole.  To account for this, we apply the correction used in \citet{Treu2007}, \citet{Bennert2010}, and \citet{Park2014}: $\log L_{V,0}=\log L_V-(0.62 \pm 0.09) \times z$, and use $L_{V,0}$ when calculating $\Delta M_{\rm{BH}}$ for Figure \ref{fig:mass_offset_evolution}.  Having made this correction, $\Delta M_{\rm{BH}}$ relative to host luminosity has a stronger redshift evolution than based on stellar mass.  We again note that the evolutionary trend is strongest among high-mass objects, and gets steeper at higher redshifts.  To compare directly with observations of \citet{Park2014} we calculate the slope of the evolution for $M_{\rm{BH}} > 10^8 M_\odot$ black holes for $z \le 0.6$, finding a best fit relation of $M_{\rm{BH}}/L_{\rm{host}} \propto (1+z)^{1.0 \pm 0.1}$, fully consistent with the observational slope of $1.2 \pm 0.7$ \citep{Park2014}.  Given this excellent agreement, we predict that the redshift evolution should get stronger at higher redshifts, which should be investigated in high-redshift surveys.

\section{Testing for sample selection effects}
\label{sec:subsampling}
The large population size of black holes in the simulation allows us
to investigate the statistical spread of subsampled populations.  To
see how the sampling method affects the results, we consider the
distribution of best-fit slopes (for all three scaling relations) at
$z=0.06$ and 1, using three different sampling methods with sample sizes
of 100 objects.  First we consider a random sampling, in which 100
black holes are selected at random from the full population.  Then we
consider populations uniformly distributed in $M_{\rm{BH}}$ ($M_*$),
which are made by taking a uniformly-distributed set of 100 masses,
ranging from $10^{6.5} M_\odot$ ($10^{10} M_\odot$) to the 15th
largest $M_{\rm{BH}}$ ($M_*$) in the population, and considering the
black hole (host galaxy) with the mass closest to the randomly
generated value.  For each sampling method, we take 10,000 samples
(randomizing the ordering of the full population for each) and find
the best fit slope and intrinsic scatter for each sample.  To avoid
undue contamination from the four strongest outliers (seen in Figure
\ref{fig:mbh_mstar}), we have removed them from consideration when
generating the subsamples (the effect of these outliers is discussed
in \S \ref{sec:subsample_outliers}).  We show the distribution of the
best-fitting slopes in Figure \ref{fig:slope_distributions} and of the
intrinsic scatter in Figure \ref{fig:scatter_distributions}.

We note several key points from Figure \ref{fig:slope_distributions}.
First, the random sample (in red) very closely matches the fit from
the full population (black vertical lines), as may be expected.  Thus
an unbiased sample of black holes will tend to provide an equivalent
slope to that of the full population.  However, the two samples that
use a uniform mass distribution (in blue and green) show significantly
steeper slopes.  As discussed in Section \ref{sec:L_dependence}, the
low-end of the relation tends to be shallower.  Since low-mass objects
are much more common than high-mass objects, the random sample tends
to be more strongly weighted toward the low-end (according to the
black hole mass function).  However, by selecting uniformly by
$M_{\rm{BH}}$ or $M_*$ we weight all mass-scales equally.  With
relatively less weight on the low-end, the resulting slope tends to be
steeper.  This sampling bias is stronger at $z=0.06$ (where the
uniform sample is 10-13 \% steeper than the random sample) than at
$z=1$ (where the uniform sample is 1-6 \% steeper than the random
sample), which can be explained by the relatively smaller range of
masses at $z=1$ than at $z=0$.  This is consistent with Figure
\ref{fig:mbh_sigma}-\ref{fig:mbh_LV}, which show the local relation to
be less linear than the $z=1$ relation.  We also note that the precise
selection method (uniform in $M_{\rm{BH}}$ or $M_*$) is unimportant,
with both methods producing equivalent results.  Rather it is when the
sample is uniformly distributed along the relation that the steeper
slope is found.

In addition to the steeper slopes, the mass-selected samples tend to
be more sharply peaked than the random sample (the standard deviation
of the random sample is roughly double that of the uniformly-selected
samples).  These narrower peaks are expected, as they have an
additional constraint imposed upon the sample selection, providing
less possibility for strongly outlying results (e.g. a sub-sample
of all low-mass objects [producing a much shallower slope] is possible
only in the random sampling).  In Table \ref{table:paramdistribution}
we provide the mean and standard devation of the distributions for the
slope and intrinsic scatter of all three scaling relations using all
three sampling methods at redshifts 0.06 and 1. We confirmed that
using smaller sample sizes than 100 (50, and 25) the distributions
broaden substantially.

We expect similar behavior in observational samples used to calculate
the slope. Observational data tend to be much more heavily weighted
toward the massive-end than a random sampling should produce (which
should follow the black hole mass function), and are roughly
equivalent to our samples uniform in mass.  Furthermore, we note that
our finding that the high-end of the relation is steeper than the low
end is consistent with observations, which have found similar behavior
in all three relations \citep{McConnellMa2012}.  The effect of sample
selection could also bias results for redshift evolution unless
carefully accounted for.  If observational samples at different
redshifts have different mass distributions (as is quite likely, due
to the difficulty observing small objects at high-$z$), then the sample
more highly-weighted toward the high-end is likely to be steeper, due
entirely to the sample selection.  We see this explicitly in Table
\ref{table:paramdistribution}, where the $M_{\rm{BH}}-\sigma$
($M_{\rm{BH}}-M_*$) relation at $z=0.06$ is 8, 15, and 17 (2, 9, 11) per
cent steeper than at $z=1$ when using random, uniform-$M_{\rm{BH}}$, and
uniform-$M_*$ distributions, respectively, despite sampling from the
same full population of black holes.

As expected, we note that the sample size has no effect on the mean slope or
scatter, and so a small sample size can still reasonably predict the
relation (Table \ref{table:paramdistribution}).  
Predictably, smaller sample sizes produce a wider distribution regardless of sampling method used, making smaller samples noticeably less reliable: a sample size of 25 results in a 1-sigma
uncertainty reaching up to 10\% of the slope, and as high as 25-30\%
of the intrinsic scatter.

\subsection{$L_{\rm{BH}}$ dependence, AGN-luminosity bias}
\label{sec:L_dependence}
\begin{figure}
\centering
\includegraphics[width=8cm]{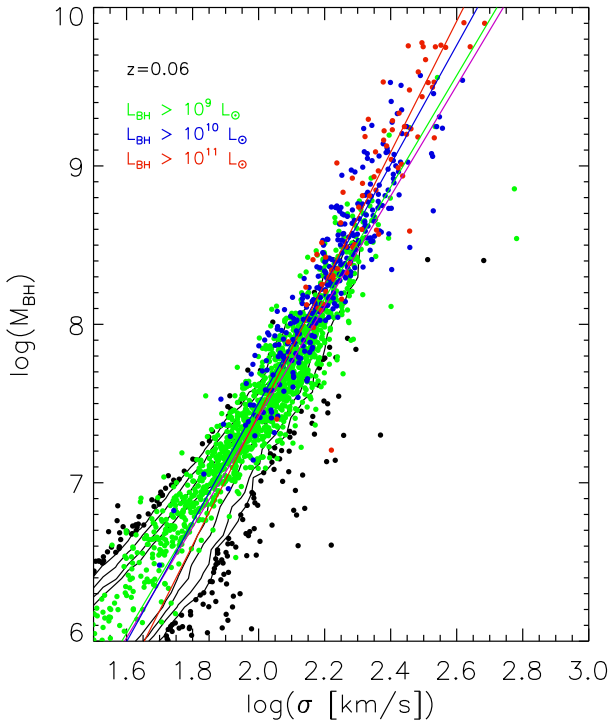}
\caption{The local $M_{\rm{BH}}-\sigma$ relation color-coded by
  lower-limits on black hole luminosity: Black - all, Green - $> 10^9
  L_\odot$, Blue - $> 10^{10} L_\odot$, Red - $> 10^{11} L_\odot$..}
\label{fig:lumdep}
\end{figure}

We depend upon the black hole luminosity ($L_{\rm{BH}}$) for many observational
studies of black holes.
Figure \ref{fig:lumdep} shows the local $M_{\rm{BH}}-\sigma$ relation,
color-coded by several lower-limits on the $L_{\rm{BH}}$ and the
associated power-law fits.  There is a clear correlation between a
black hole's mass and luminosity, with brighter black holes tending to
be larger mass.  This is expected, given that the accretion rate depends on
$M_{\rm{BH}}^2$ (or $M_{\rm{BH}}$, if Eddington limited).  However, we
note that for a given blackhole mass we tend to find a range of
luminosities, as has been previously shown \citep{DeGrafBHgrowth2012}.
This range in luminosities means that even when using a
luminosity-limited sample, we obtain a similar result to using the
full sample.  We demonstrate this in Table \ref{table:paramdep}, where
the best-fitting parameters for all three scaling relations are listed
for cuts on $L_{\rm{BH}}$.  In all three relations
($M_{\rm{BH}}-\sigma$, $M_{\rm{BH}}-M_*$, and $M_{\rm{BH}}-L_V$) we
see the same qualitative behavior: a higher luminosity threshold tends
to be slightly larger (higher $\alpha$), has a steeper dependence on
the host (higher $\beta$), and a larger intrinsic scatter (larger
$\epsilon_0$).  However, we note that this dependency is very weak.
The normalization ($\alpha$) only increase by $\sim 0.2$ dex going
from $L_{\rm{BH}} > 10^8 L_\odot$ to $> 10^{11} L_\odot$.  The slope
increase by $\sim 6-10\%$ between cuts of $10^8 L_\odot$ and $10^{10}
L_\odot$, and although the cut of $10^{11} L_\odot$ may be a bit
steeper, the sample size is much smaller with such a high cut (78
objects).

However, when we perform a test based on a faint sample ($L_{\rm{BH}}
< 10^{11} L_\odot$), a bright sample ($L_{\rm{BH}} > 10^{11}
L_\odot$), and a subsample of faint black holes matched
to the bright sample according to host stellar mass we find a significant
difference.  Because the most massive faint black hole is only
$10^{9.57} M_\odot$, we only consider BHs below this mass for this
sample comparison.  With this upper limit on $M_{\rm{BH}}$, we find
the bright sample slope to be 3.93, and the faint sample slope to be
3.38.  To generate the matched sample, for each bright black hole we
find the black hole from the faint sample which most closesly matches 
it in host stellar mass ($M_*$).
Repeating this process for 1000 randomized
orderings, we find the $M_*$-matched sample to have a slope of $3.89
\pm 0.02$, fully consistent with the bright sample.  Thus we find that
although the correlations for bright AGNs may appear to be steeper than fainter
ones, it is due entirely to the different distribution of host masses that the bright AGN populate.
Specifically,
bright black holes tend to be more massive black holes and located in larger hosts than fainter ones, but
there is no inherent difference between bright and faint AGN
of equivalent black hole mass/within equivalent host masses. 

Thus there is a bias when black holes are luminosity-selected without any additional
considerations, which must be accounted for in flux-limited observations (particularly at
high-redshift where the luminosity limits are higher).  Any such flux-limited survey must take 
the $M_{\rm{BH}}$ distribution into consideration, and should only be compared to measurements from similar $M_{\rm{BH}}$-distribution samples (or else recognize that the higher flux limit will bias the result toward a steeper slope).  This is particularly important for evolution studies, where a fixed flux limit will bias the high-redshift observations toward a steeper slope unless the mass distribution is accounted for.

Similar to the $L_{\rm{BH}}$-dependence, we consider the dependence on AGN-activity.  Using a cut on Eddington fraction ($f_{\rm{edd}}= \dot{M}_{\rm{BH}}/\dot{M}_{\rm{edd}}$) as the theshold for activity, we find weak evidence for a dependence on activity at low redshift.  At $z=0.06$, the best fitting $M_{\rm{BH}}-\sigma$ slope for BHs with $f_{\rm{edd}} > 0.1$ (0.05) is 5.6 (5.0), compared to the full-sample slope of 3.5.  As done with the $L_{\rm{BH}}$ sample, we create a subsample of inactive BHs mass-matched (using $M_*$) to the active sample, providing slopes of $3.5 \pm 1$ ($3.6 \pm 0.7$).  As $z=0.3$ we find an active slope of 4.83 (4.1), while the mass-matched inactive samples have slopes of $3.8 \pm 1.4$ ($3.5 \pm 0.5$) for $f_{\rm{edd}} > 0.1$ (0.05), while for $z > 0.3$ we find no difference between the active and inactive populations.  Thus we find the local relation is steeper for active black holes, but only slightly outside the 1-$\sigma$ variation of a comparable inactive sample.

\subsection{High-mass sample}
\label{sec:highmass}

To consider the relation for high-mass black holes, in Table \ref{table:highmassparams} we provide the best fitting parameters using only black holes above $10^8 M_\odot$ for $z \le 2$ (at $z=4$ we don't have enough large black holes for an accurate fitting).  To confirm that using a strict cut on $M_{\rm{BH}}$ doesn't bias our result, we also test using a cut perpendicular to the full relation at $10^8 M_\odot$.  The slope and normalization are essentially unaffected (less than 5\% difference between selection methods).  However, using a strict cut on $M_{\rm{BH}}$ results in a smaller intrinsic scatter, so we provide the intrinsic scatter for both selection methods.  

Compared to the full relation, we find several important differences in the high-mass sample.  First, the high-mass sample is generally $\sim$0.2 dex higher than the full sample, suggesting that massive black holes tend to be about 50\% larger relative to their host galaxy than low-mass black holes.  At low redshift we find a larger intrinsic scatter among the high-mass sample, due in large part to the strongly undermassive outliers at low-$z$.  At higher redshift (redshift above $\sim 1$) the trend reverses, with the high-mass sample (where we have fewer outliers) having a smaller scatter than the full relation.  This can also be seen in Figures \ref{fig:mbh_sigma}-\ref{fig:mbh_LV}, where the largest scatter in the relation tends to be at moderate masses.  

Most significantly, however, we find the high-mass slope to be shallower than that of the full sample \citep[consistent with ][]{Volonteri2011}, and the discrepancy becomes more significant at higher redshift.  At $z=1$, the high mass samples are 20\%, 29\%, and 53\% steeper than the full sample for $M_{\rm{BH}}-\sigma$, $M_{\rm{BH}}-M_*$, and $M_{\rm{BH}}-L_V$, respectively.  This shallower slope at high mass suggests that once black holes reach a sufficiently high mass, they begin growing slower relative to their host galaxy, fully consistent with previous work which found that high-mass black holes tend to grow slower than moderate mass black holes \citep{DeGrafBHgrowth2012}.  This behavior is investigated in more detail in Section \ref{sec:tracing_evolution}.  The shallower slope is also very important for high-redshift observations, which are generally only able to constrain the properties of high-mass black holes and thus might underestimate the slope of the full sample if the selection function is not properly accounted for.

\begin{table}
\centering
\begin{tabular}{c c c c}
& && \\
\multicolumn{4}{c}{Parameters for relation using high-mass sample} \\

\hline
\hline
 $z$  & $\alpha$ & $\beta$ & $\epsilon_0$ \\
\hline
\multicolumn{4}{c}{$M_{\rm{BH}}-\sigma$ relation ($x_0 = 200 \rm{km/s}$)}\\
\hline
 0.06 & $8.582 \pm 0.011$ & $3.270 \pm 0.094$ & $0.225$ (0.221) \\
 0.3 & $8.434 \pm 0.010$ & $3.073 \pm 0.093$ & $0.189$  (0.190)\\
 0.6 & $8.327 \pm 0.010$ & $2.812 \pm 0.092$ & $0.182$  (0.180)\\
 1 & $8.285 \pm 0.009$ & $2.706 \pm 0.087$ & $0.166$  (0.183)\\
 2 & $8.301 \pm 0.012$ & $2.315 \pm 0.125$ & $0.178$  (0.184)\\

\hline
\multicolumn{4}{c}{$M_{\rm{BH}}-M_*$ relation ($x_0 = 10^{11} M_\odot$)}\\
\hline
 0.06 & $8.122 \pm 0.015$ & $1.102 \pm 0.032$ & $0.239$ (0.249) \\
 0.3 & $8.073 \pm 0.014$ & $1.039 \pm 0.030$ & $0.191$  (0.207)\\
 0.6 & $8.046 \pm 0.013$ & $0.978 \pm 0.030$ & $0.182$  (0.191)\\
 1 & $8.080 \pm 0.013$ & $0.942 \pm 0.031$ & $0.176$  (0.197)\\
 2 & $8.250 \pm 0.013$ & $0.843 \pm 0.043$ & $0.180$  (0.217)\\

\hline
\multicolumn{4}{c}{$M_{\rm{BH}}-L_V$ relation  ($x_0 = 10^{10.5} L_\odot$)}\\
\hline
 0.06 & $8.468 \pm 0.014$ & $0.825 \pm 0.036$ & $0.308$ (0.354) \\
 0.3 & $8.389 \pm 0.011$ & $0.848 \pm 0.033$ & $0.237$  (0.283)\\
 0.6 & $8.287 \pm 0.011$ & $0.793 \pm 0.033$ & $0.221$  (0.247)\\
 1 & $8.185 \pm 0.013$ & $0.782 \pm 0.035$ & $0.215$  (0.254)\\
 2 & $8.117 \pm 0.021$ & $0.625 \pm 0.043$ & $0.212$  (0.259)\\

\end{tabular}
\caption{Best fitting parameters for the scaling relations (Equation
  \ref{eqn:fit}) based on a sample of black holes with $M_{\rm{BH}} > 10^8 M_\odot$.  We also provide $\epsilon_0$ for a cut perpendicular to the full relation in parentheses (see text).} 
\label{table:highmassparams}
\end{table}

\subsection{Intrinsic scatter}
\label{sec:scatter}

\begin{figure}
\centering
\includegraphics[width=9cm]{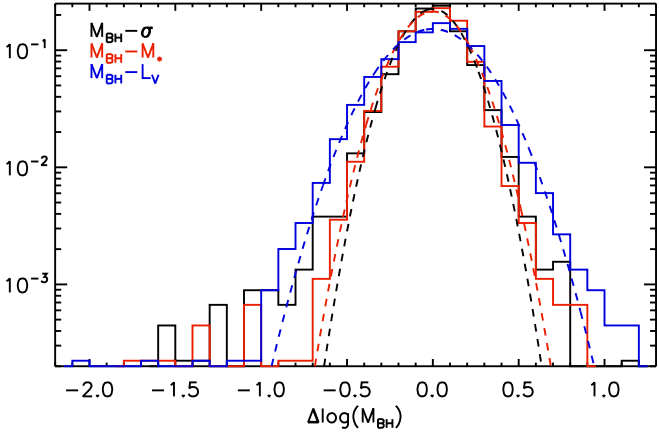}
\caption{Distribution of black hole offset relative to the best fitting scaling relations provided in Table \ref{table:mainparams}: 
\textit{Black:} $M_{\rm{BH}}$ offset relative to $M_{\rm{BH}}-\sigma$ relation
\textit{Red:} $M_{\rm{BH}}$ offset relative to $M_{\rm{BH}}-M_*$ relation
\textit{Blue:} $M_{\rm{BH}}$ offset relative to $M_{\rm{BH}}-L_V$ relation.  Dashed lines show the best-fitting Gaussian distributions.}
\label{fig:scatter}
\end{figure}

Having found the best fitting functions for the main scaling relations, we also investigate the distribution of black holes about this relation.  In Figure \ref{fig:scatter} we plot the distribution function of $\Delta \log (M_{\rm{BH}})$, the offset of the black hole mass from each of the best-fitting scaling relations (solid histograms: black - $M_{\rm{BH}}-\sigma$; red - $M_{\rm{BH}}-M_*$; blue - $M_{\rm{BH}}-L_V$) at redshift zero.  We find that the scatter in the scaling relations tends to be well fit by Gaussian distributions, showing the distribution is generally symmetric about the mean relation (except for low-end outliers, discussed below, and in Section \ref{sec:subsample_outliers}).  For each relation, we overplot the Gaussian with standard deviation equal to the intrinsic scatter found in Section \ref{sec:Results}.  The good agreement shows that the distribution of black holes about the best fitting relation is indeed Gaussian, and that the intrinsic scatter $\epsilon_0$ can be well-calculated as in Section \ref{sec:msigma}.  The Gaussian nature of the scatter in each of the scaling planes is particularly important for observational studies, where knowledge of the underlying scatter is needed for creating a selection function.

\begin{figure}
\centering
\includegraphics[width=9cm]{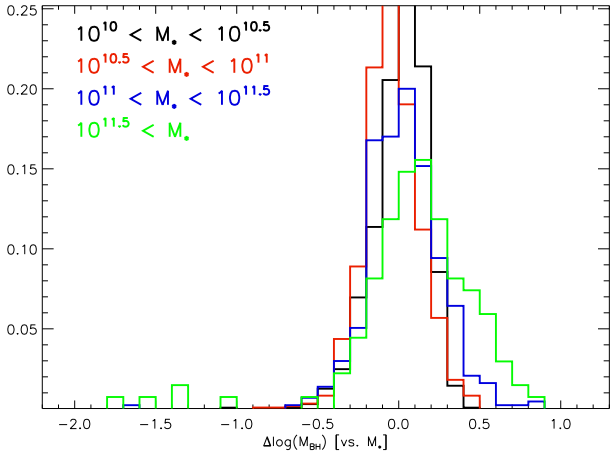}
\caption{Distribution of black hole offset relative to $M_{\rm{BH}}-M_*$ relation at $z=0.06$, binned by host stellar mass.  Black - $10^{10} < M_* < 10^{10.5}$, Red - $10^{10.5} < M_* < 10^{11}$, Blue - $10^{11} < M_* < 10^{11.5}$, Green - $10^{11.5} < M_*$.}
\label{fig:scatter_massdep}
\end{figure}

In Figure \ref{fig:scatter_massdep} we show the distribution of black hole offsets based on the $M_{\rm{BH}}-M_*$ relation, binned by host stellar mass.  For low- to moderate-stellar mass, the distribution is roughly $M_*$-independent.  At the highest masses ($M_* > 10^{11.5} M_\odot$), however, the distribution is shifted upward, such that the black holes tend to lie above the relation by $\sim$0.1 dex, with a larger high-end tail to the distribution.  We also find that the majority of the significantly undermassive objects (at least 1 dex below mean relation) tend to be found in these very high-mass hosts, and can be identified as they are strong outliers from the otherwise Gaussian distribution.

\subsection{Outliers}
\label{sec:subsample_outliers}

\begin{figure}
\centering
\includegraphics[width=9cm]{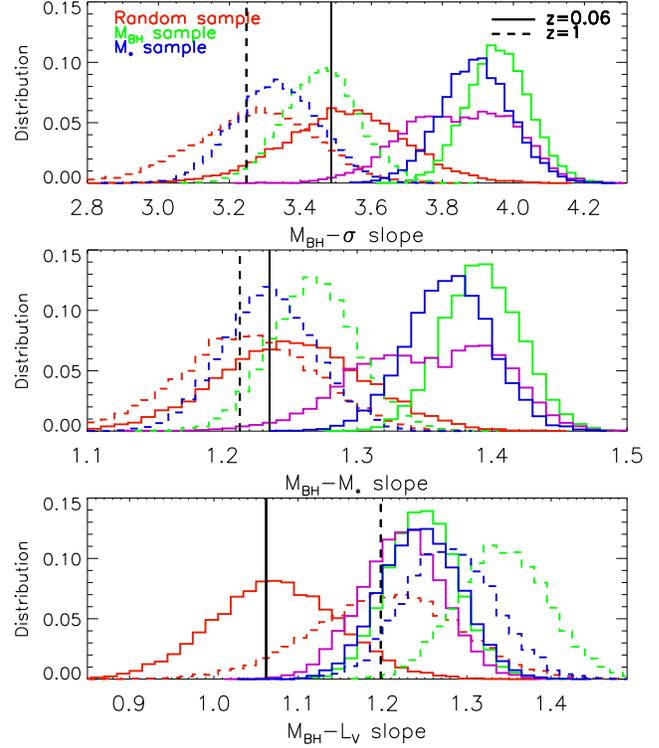}
\caption{The probability distribution function of the best fit slopes
  of $M_{\rm{BH}}-\sigma$ (top), $M_{\rm{BH}}-M_*$ (middle), and
  $M_{\rm{BH}}-L_V$ (bottom) using subsamples of 100 objects from our
  full population at $z=0.06$ (solid lines) and $z=1$ (dashed lines).  The
  sampling techniques are a random sampling (red), a sample uniformly
  distributed in $M_{\rm{BH}}$ (green), and a sample uniformly
  distributed in $M_*$ (blue).  Each distribution is found using
  10,000 distinct samplings, after removing the four strongest
  outliers from the $z=0$ population (see text).  In pink we show the
  distribution of the $M_{\rm{BH}}$-selected sample if those outliers
  are not removed.  We also show the best fit slope from the full
  population as vertical lines.}
\label{fig:slope_distributions}
\end{figure}

\begin{figure}
\centering
\includegraphics[width=9cm]{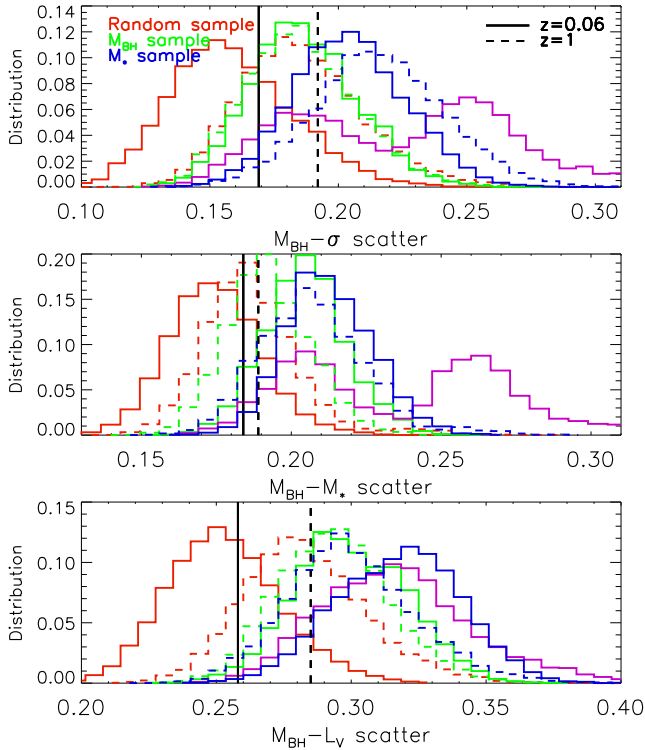}
\caption{As Figure \ref{fig:slope_distributions} but for the intrinsic scatter ($\epsilon_0$) rather than the slope.}
\label{fig:scatter_distributions}
\end{figure}

As mentioned in Section \ref{sec:subsampling}, the distributions in
Figures \ref{fig:slope_distributions} and
\ref{fig:scatter_distributions} are based on samples which neglect the
strongest outliers, identifiable as being well outside the otherwise Gaussian scatter about the mean relation (Section \ref{sec:scatter}).  Including those outliers has minimal effect on
the random distribution (since they are so rarely selected), however
the uniform distributions include them much more frequently, resulting
in bimodal distributions.  We show this in Figures
\ref{fig:slope_distributions} and \ref{fig:scatter_distributions} with
a pink curve, which uses uniform samples in $M_{\rm{BH}}$ based on all
BHs (i.e. equivalent to the green curve except with the outliers
included).  In the $M_{\rm{BH}}-L_V$ relation, the inclusion of these
objects has minimal effect, since that relation's larger intrinsic
scatter means they are much weaker outliers.  In the
$M_{\rm{BH}}-\sigma$ and $M_{\rm{BH}}-M_*$ relations, however, we can
see a strong change when including these outliers, resulting in a much
broader distribution of slopes (standard deviation of 0.15 compared to
0.09 when the outliers are removed) and a clear bimodality in the
distribution of intrinsic scatter (samples without any outliers have
$\epsilon_0 \sim 0.19$ compared to $\epsilon_0 \sim 0.25$ for samples
which include an outlier).

When fitting these subsamples, we found an interesting bimodality
arising from the four outlying points seen in the $z=0$ panel of Figure
\ref{fig:mbh_mstar}.  In these smaller samples, the inclusion of one
of these outlying points had a significant effect on the slope and
especially the intrinsic scatter, produing a wide distribution of
slopes and a clearly bimodal distribution of the scatter.  To avoid
such strong contamination from only a few objects, we remove those
four outliers and re-calculate the fits, resulting in the elimination of the bimodality.  We note that the size of the secondary peak caused by the inclusion of at least one outlier is strongly dependent on the sample size used.  Smaller sample sizes have a smaller secondary peak (since it is less likely to capture one of the outliers), but the secondary peak moves further from the primary peak (since the effect of the outlier is larger in a smaller sample).

\begin{table}
\centering
\begin{tabular}{c c c c c c}

\multicolumn{6}{c}{Distribution of fitted parameters} \\

\hline
\hline

$z$ & Selection method & $< \beta >$ & $\sigma (\beta)$ & $< \epsilon_0 >$ & $\sigma (\epsilon_0)$ \\

\hline
\multicolumn{6}{c}{$M_{\rm{BH}}-\sigma$}\\
\hline
0.06 & Random & 3.52 & 0.17 & 0.16 & 0.02 \\
     & Uniform $M_{\rm{BH}}$ & 3.96 & 0.09 & 0.19 & 0.02 \\
     & Uniform $M_{*}$ & 3.89 & 0.10 &  0.20 & 0.02 \\
1 & Random & 3.27 & 0.17 & 0.19 & 0.02 \\
     & Uniform $M_{\rm{BH}}$ & 3.46 & 0.11 & 0.19 & 0.02 \\
     & Uniform $M_{*}$ & 3.33 & 0.12 & 0.22 & 0.03 \\

\hline
\multicolumn{6}{c}{$M_{\rm{BH}}-M_*$}\\
\hline

0.06 & Random & 1.25 & 0.06 & 0.17 & 0.02 \\
     & Uniform $M_{\rm{BH}}$ & 1.39 & 0.03 & 0.20 & 0.01 \\
     & Uniform $M_{*}$ & 1.37 & 0.03 & 0.21 & 0.01 \\

1 & Random & 1.22 & 0.05 & 0.19 & 0.02\\
     & Uniform $M_{\rm{BH}}$ & 1.27 & 0.03 & 0.19 & 0.01 \\
     & Uniform $M_{*}$ & 1.23 & 0.03 & 0.21 & 0.02 \\

\hline
\multicolumn{6}{c}{$M_{\rm{BH}}-L_V$}\\
\hline

0.06 & Random & 1.07 & 0.07 & 0.25 & 0.02 \\
     & Uniform $M_{\rm{BH}}$ & 1.25 & 0.04 & 0.30 & 0.02 \\
     & Uniform $M_{*}$ & 1.25 & 0.05 & 0.32 & 0.02 \\
1 & Random & 1.20 & 0.09 & 0.28 & 0.02 \\
     & Uniform $M_{\rm{BH}}$ & 1.34 & 0.05 & 0.29 & 0.02 \\
     & Uniform $M_{*}$ & 1.28 & 0.05 & 0.30 & 0.02 \\

\end{tabular}
\caption{Mean and standard deviation of slope ($\beta$) and intrinsic scatter ($\epsilon_0$) for subsamples of 100 objects selected at random, or with uniform distributions in $M_{\rm{BH}}$ or $M_*$ (see text for details).  }
\label{table:paramdistribution}
\end{table}

\section{Evolution of BH holes on the $M_{\rm{BH}}-M_*$
  plane}\label{sec:tracing_evolution}

\begin{figure*}
\centering
\includegraphics[width=15cm]{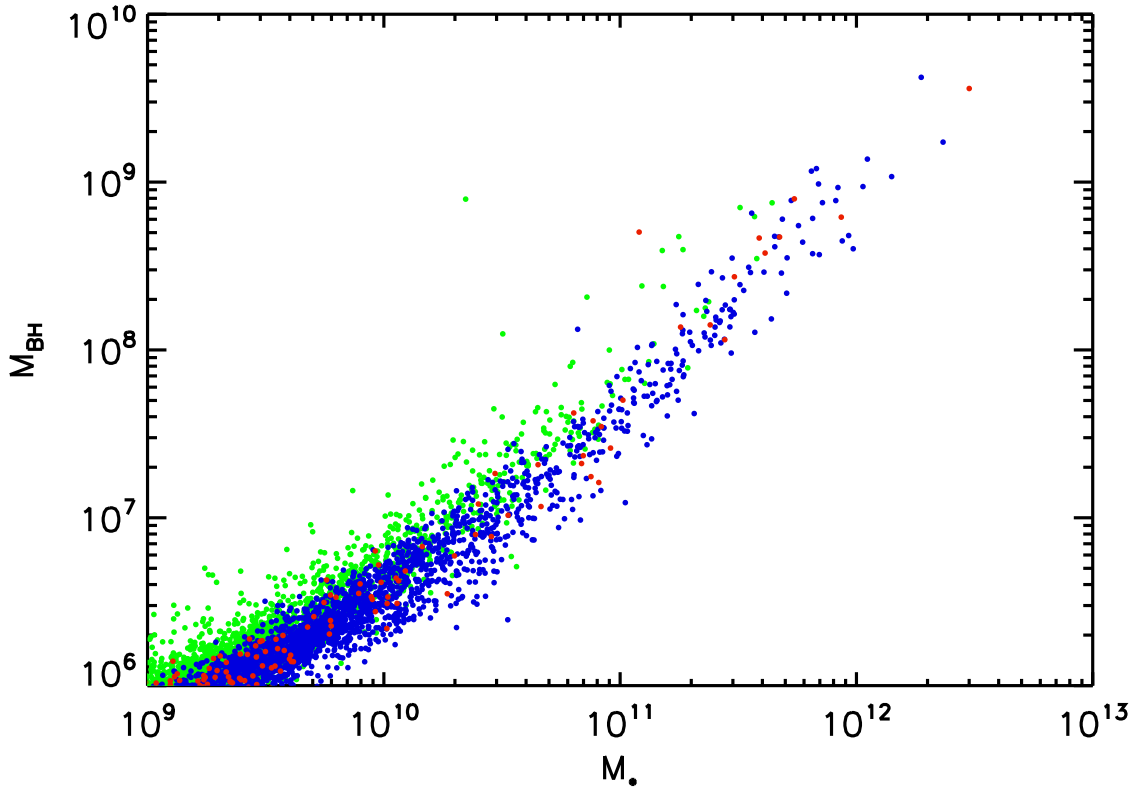}
\caption{$M_{\rm{BH}}-M_*$ plot at $z=0.6$, color-coded by individual
  black hole evolution along the $M_{\rm{BH}}-M_*$ plane.  The BH
  population is divided into three categories based on evolution from
  $z=0.6$ to $z=0.06$: Green - slope along $M_{BH} - M_*$ plane is
  positive, but shallower than the local relation; Blue - slope is
  positive, and steeper than the local relation; Red - slope is
  negative (i.e. rare case where $M_*$ has decreased).  To avoid
  spurious slopes from small evolution, only objects with a change of
  at least 30\% in both $M_{BH}$ and $M_*$ are included.}
\label{fig:evolution_along_plane}
\end{figure*}

In Figure \ref{fig:evolution_along_plane} we show the
$M_{\rm{BH}}-M_*$ relation at $z=0.6$, with each black hole color-coded
by the slope of its evolution along the $M_{\rm{BH}}-M_*$ plane from
$z=0.6$ to $z=0.06$.  Green points show black holes which evolve along a
shallower slope than the best fit local relation (i.e. black holes
which are becoming less massive relative to their host).  Blue points
show black holes which are steeper than the local relation (i.e. black
holes which are becoming more massive relative to their host).  The
red points are cases in which the slope is negative, due to the
stellar mass of the host decreasing from $z=0.6$ to $z=0.06$.  Ignoring
the rare cases where the host galaxy has gotten smaller, we find a
clear trend in the slopes displayed here.  Black holes which are
overmassive for their given host mass tend to grow slower than there
host (i.e. shallower slope than the local relation, so blue points),
bringing them toward the local relation.  In contrast, the
undermassive black holes tend to grow more rapidly, bringing them up
toward (and often above) the relation.  

\begin{figure}
\centering
\includegraphics[width=9cm]{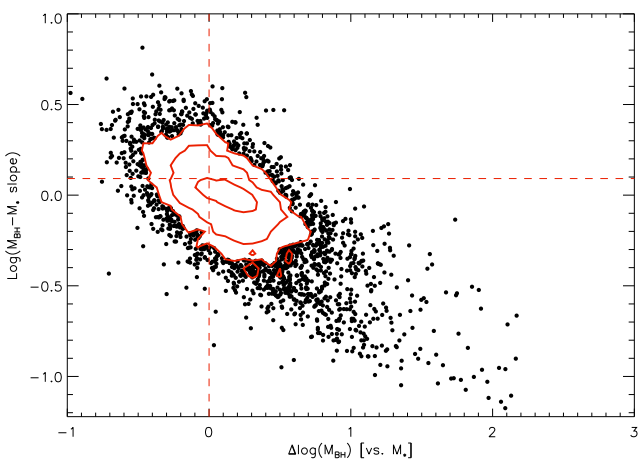}
\caption{Correlation between the black hole offset relative to the local $M_{\rm{BH}}-M_*$ relation ($\Delta \log (M_{\rm{BH}})$) and the slope of its trajectory along the $M_{\rm{BH}}-M_*$ plane.  The vertical dashed line is the threshold for lying on the local $M_{\rm{BH}}-M_*$ plane, and the horizontal dashed line is the slope of the local relation.}
\label{fig:slope_vs_offset}
\end{figure}

This is shown statistically in Figure \ref{fig:slope_vs_offset}, which plots the correlation between $\Delta M_{\rm{BH}}$ (i.e. how overmassive the black hole is relative the host stellar mass) and the slope of the black holes trajectory along the $M_{\rm{BH}}-M_*$ plane from $z=0.3$ to 0.06.  Here we clearly see that the more overmassive the black hole, the shallower its trajectory along the scaling relation.  Overmassive black holes ($\Delta \log (M_{\rm{BH}}) > 0$) are much more likely to move on a shallow trajectory, with 86\% having a shallower trajectory than the slope of the local relation.  49\% of undermassive black holes have a shallower trajectory, falling to 28\% for black holes undermassive by at least 0.3 dex.  The best fit relation between the slope trajectory and $\Delta \log (M_{\rm{BH}})$ is $1.08 \times (\Delta \log (M_{\rm{BH}}))^{-0.46}$.  This is consistent with our
general picture for how black holes evolve with respect to their
galaxies: low-mass black holes tend to grow quickly, bringing them up
and often above the local relation; upon reaching a sufficiently large
mass (i.e. above the relation), the black hole enters a self-regulated
regime, suppressing its further growth while the host continues to
grow, bringing it back to the general relation we find.  This is also supported by the $M_{\rm{BH}}$-dependence of the relation (Table \ref{table:highmassparams}), which found the high-mass black holes (which tend to be overmassive and in the self-regulated regime) have a shallower slope than lower-mass objects.  We note that \citet{Volonteri2009} find similar behavior for low-mass black hole growth fueled by galaxy mergers, where black holes seeded below the relation tend to grow rapidly without significant self-regulation, while the more massive objets tend to have minimal growth until the host halo has grown significantly.

\begin{figure}
\centering
\includegraphics[width=9cm]{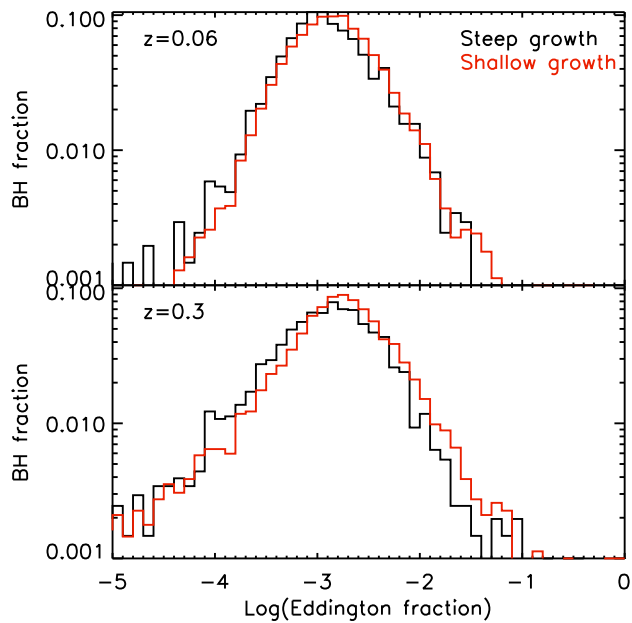}
\caption{Distribution of the Eddington fraction ($\dot{L}_{\rm{BH}}/ \dot{L}_{\rm{edd}}$) at $z=0.06$ (top) and $z=0.3$ (bottom) for black holes whose evolution along the $M_{\rm{BH}}-M_*$ plane steeper (black) or shallower (red) than the slope of the local relation. }
\label{fig:active_distribution}
\end{figure}

To determine if the trajectory a black hole takes along the scaling relation is a function of its activity, in Figure \ref{fig:active_distribution} we show a histogram of the Eddington fraction of the black holes at $z=0.06$ (top) and 0.3 (bottom), divided into two populations: one for black holes whose path on the $M_{\rm{BH}}-M_*$ plane is steeper than the local relation (black) and one for those whose path is shallower (red).  We find that there is minimal difference between the two distributions, suggesting that the path taken along the scaling plane is not strongly dependent on the black holes' Eddington fraction.  However, we note that this is based upon the slope for a single step from $z=0.3$ to $z=0.06$, while the accretion rate of the black hole (and thus the Eddington fraction) varies on much shorter timescales.  
To fully characterize the correlation (if any) between AGN activity and motion along the scaling plane will require an investigation into the co-evolution of BH and host galaxy with much finer time resolution, which we leave for a future work.

\begin{figure*}
\centering
\includegraphics[width=15cm]{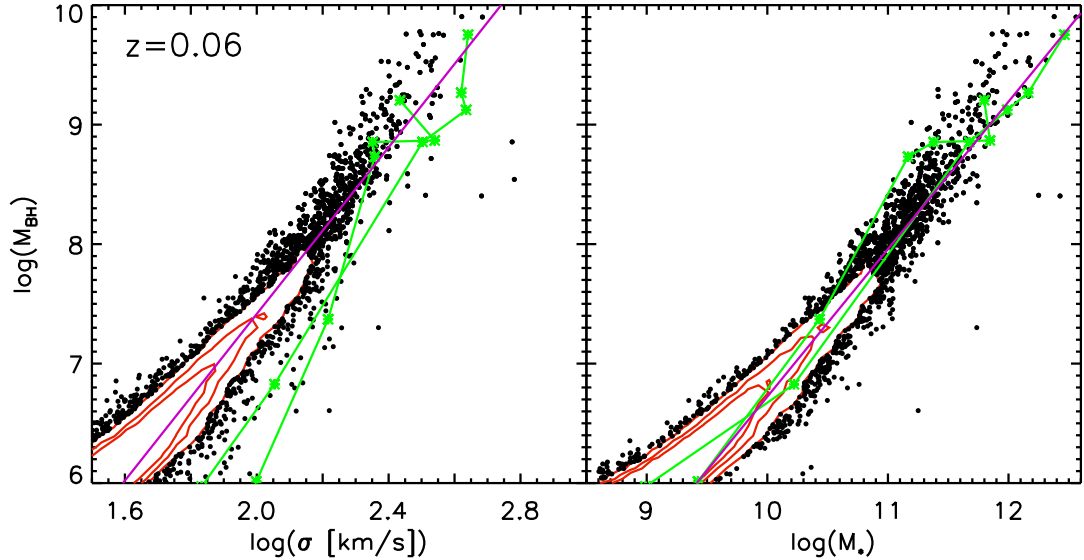}
\caption{The $M_{\rm{BH}}-\sigma$ and $M_{\rm{BH}}-M_*$ (right) relations at $z=0.06$, with tracks from two individual black holes shown in green.  These tracks show the evolution of each object over their lifespan, until reaching their final location at the end of the simulation.  }
\label{fig:individual_tracks}
\end{figure*}

To explicitly illustrate black hole evolution along the scaling relations, in Figure \ref{fig:individual_tracks}
we show the $M_{\rm{BH}}-\sigma$ and $M_{\rm{BH}}-M_*$ planes at
$z=0.06$, with the tracks of two black holes overplotted on top in
green, and the local relation shown in pink.  These two tracks show
the typical behavior described here.  In each case, the black hole
grows along a trajectory steeper than the local relation, bringing it
up to the local relation \citep[similar to][]{Volonteri2009}.  This continues until the end of the
simulation for the more massive BH, without reaching a regulated
regime (but also without rising above the local relation).  The second
BH, however, does reach a self-regulated regime.  In the
$M_{\rm{BH}}-\sigma$ plane it only barely surpasses the local
relation, but it more strongly overcomes it in the $M_{\rm{BH}}-M_*$
plane.  At this point the black hole growth slows substantially, while
the host continues growing (in both $\sigma$ and $M_*$), bringing the
black hole under the local relation.  Once below the local relation,
the host galaxy undergoes a merger.  This merger is sufficient to
restart the black hole growth, bringing the black hole back to
slightly above the local relation.

\section{Conclusions}
\label{sec:Conclusions}

We have presented the scaling relation between supermassive black
holes and their host galaxies from the high-resolution cosmological
simulation \textit{MassiveBlackII}.  The volume of this simulation
provides a large sample size of black holes reaches $\sim 10^{10}
M_\odot$ by $z \sim 0$, which we use to study the scaling relations
across a wide range of redshifts, considering both how the relation
evolves with cosmic time, and how individual black holes tend to
evolve relative to the relation.

The simulation does a good job reproducing all three major scaling
relations investigated: $M_{\rm{BH}}-\sigma$, $M_{\rm{BH}}-M_*$, and
$M_{\rm{BH}}-L_V$.  In the $M_{\rm{BH}}-\sigma$ relation, we match the
high-end data very closesly.  The low-end of our results tend to lie
above the observations, but given the much closer agreement in the
relations with $M_*$ and $L_V$, we expect this to be primarily due to
the computation of $\sigma$ rather than an overestimate in the black
hole mass.  This is consistent with \citet{Sijacki2014}, who found
similar behavior at the low-end, at least in part due to the
contributions from a non-negligible rotational component to $\sigma$.
This will be addressed in a future work on the full
bulge-decomposition of galaxies within the simulation.

The $M_{\rm{BH}}-M_*$ and $M_{\rm{BH}}-L_V$ relations, both based on
the total galaxy rather than bulge properties, also show good
agreement at the high end, as well as closer agreement at the low-end
of the relation, particularly in $M_*$, the more fundamental of the
two properties within our simulation.  The $M_{\rm{BH}}-L_V$ relation
has the largest scatter among the three relations investigated,
suggesting it to be a less fundamental relation than either of the
other two.  This lesser correlation is not surprising, given that the
stellar mass is a more fundamental quantity during the run, and the
velocity dispersion is more directly related to the potential well of
the host.  The low-end $M_{\rm{BH}}-L_V$ relation tends to flatten out
at high-$z$ as the black hole seed mass is approached, showing a clear
break in the relation.  However, this is in the least well-resolved
objects (both quite small and quite young), and given that the break
disappears after $z \sim 2$ and isn't found in either $\sigma$ or
$M_*$, we expect this is likely an artifact of the simulation.

Within each of the three scaling relations, we test the scatter of black holes above and below the local relation, and find it to be well-described by a Gaussian distribution (with standard deviation equal to the intrinsic scatter $\epsilon_0$ in Table \ref{table:mainparams}), except in the highest mass hosts.  In the largest host galaxies ($M_* > 10^11.5 M_\odot$) the scatter is biased toward larger (by $\sim$0.1 dex) black holes, with a stronger high-end tail to the distribution.  Thus we suggest that observations should not assume a Gaussian spread in the selection function among high-mass objects.  

We consider the evolution of the relation with redshift, and find none
of the three relations exhibit strong evolution for $z \le 1$.
Above redshift 1, black holes of a given mass tend to be found in
higher-$\sigma$, lower-$L_V$ hosts.  This evolution is found to be
mass-dependent, with massive black holes being found in slightly
less-massive hosts (shown explicitly as a mass-offset relative to the
local relation).  This is consistent with previous findings that
massive black hole growth is more rapid at high-redshifts
\citep{DeGrafBHgrowth2012}.

In contrast to the offset evolution based on $M_*$, the offset
relative to $L_V$ decreases with time, suggesting the black holes of a
given mass tend to be found in brighter hosts at high-$z$.  This
evolution is due primarily to the evolution of the mass-to-light ratio
(with high-$z$ galaxies tending to be brighter than their low-$z$
counterparts), rather than an evolution in the BH-host relation. This
is clearly seen when compared to the $M_{\rm{BH}}-M_*$ relation, which
shows the opposite trend, or when considering the correlation between $M_{\rm{BH}}$ and $L_{V,0}$, i.e. the luminosity corrected for passive evolution.  
To compare directly with observations, we consider the evolution in offset based on the $M_{\rm{BH}}-L_{V,0}$ local relation using only black holes with $M_{\rm{BH}} > 10^8 M_\odot$ and $z \le 0.6$, finding a relation of $M_{\rm{BH}}/L_{V,0} \propto (1+z)^{1.0 \pm 0.1}$. This is fully consistent with the current observational measurement of $M_{\rm{BH}}/L_{host,0} \propto (1+z)^{1.2 \pm 0.7}$ \citep[][ private correspondence]{Park2014}.

When considering cuts on the black hole luminosity, we find all three
relations to be nearly completely $L_{\rm{BH}}$-independent.  From a
theoretical standpoint, this is significant as it demonstrates that
black holes in any given region of the BH-host plane can span a wide
range of accretion rates (though with a general trend of larger black
holes tending to be brighter).  In other words, we expect any given
set of $M_{\rm{BH}}-\sigma, M_*, L_V$ values to include some black
holes accreting at unusually high eddington fractions, while others
will be at unusually low fractions.  From an observational standpoint,
this is significant since it means that for any given survey, the flux
limit imposed by the instrumentation will limit only the sample size
that can be measured and the range over which the sample will span;
the flux limits will \textit{not} result in a bias for the detected
slope of the relation being studied, which will be of crucial
importance for upcoming high-redshift surveys.

In addition we consider the relation of only the high-mass black holes, finding they tend to be $\sim$0.2 dex larger relative to their host.  However, the high-mass sample has a shallower slope, suggesting that the highest-mass black holes grow slower (relative to their host galaxy) than more moderate mass objects.  
We investigate this in more detail by considering the behavior of individual black holes and
their evolution along the BH-host planes.  In general, we find that
black holes tend to evolve toward the local relation, resulting in a
general decrease in the intrinsic scatter within the relation.  Black
holes below the local relation tend to grow more rapidly, bringing
them up toward (and even above) the general relation.  Once grown to
be `overmassive' relative to the relation, a self-regulated regime may
be entered, suppressing further black hole growth without necessarily
inhibiting host growth.  This results in a much shallower evolution
along the plane, bringing the overmassive black hole back toward the
mean relation.  Thus we generalize the black hole behavior as initial
rapid growth, suppressed only upon surpassing the mean relation, at
which point regulation of the black hole growth brings it back toward
the typical relation.

Finally, we consider the effect sampling has on the scaling relation.
We find that truly random sampling is equivalent to the full
population, though with substantial spread in the distribution,
especially for smaller sample sizes (e.g. standard deviation in the
$M_{\rm{BH}}-\sigma$ slope is $\sim$0.17 for a sample of 100, and
$\sim$0.35 for a sample of 25).  However, both the full population and
the random sub-samples tend to be dominated by the low end (since the
low-mass objects drastically outnumber the massive ones).  In contrast
to this, a sample with a uniform distribution of masses (either
$M_{\rm{BH}}$ or $M_*$) will not be affected this way.  By probing all
scales equally, the result is less biased toward the typically
shallower low-end, producing a much steeper result.  Furthermore, the
distribution of slopes tends to have a narrower distribution, since the uniform distribution in mass makes a biased subsample much less likely.
We also note that distributions which uniformly span the mass range
considered tend to find a stronger redshift evolution than a random
sampling of objects does.  Since the high-mass objects tend to evolve
more quickly, the sampling which weights the high-end more strongly
(the uniform distribution) will necessary find a quicker evolution
than one which weights the low-end more strongly (the full population
and the random sampling).  This possible bias toward
redshift-evolution of the slope of the relations must be carefully
addressed in any high-redshift observational study.

\section*{Acknowledgments}

The simulations were run on NSF XSEDE facilities (Kraken
  at the National Institute for Computational Sciences). We
  acknowledge support from Moore foundation which enabled us to
  perform the simulations and data analysis at the McWilliams Center
  of Cosmology at Carnegie Mellon University.  TDM has been funded by
  the National Science Foundation (NSF) PetaApps, OCI-0749212 and by
  NSF AST-1009781 and ACI-1036211.
TT acknowledges support from the Packard Foundation through a Packard Research Fellowship, from NASA through grants HST-AR-12625, GO-10216, GO-11208, GO-11166.
We thank Matthew Auger and Vardha Bennert for comments and discussion.

 \bibliographystyle{mn2e}       
 \bibliography{astrobibl}       

\end{document}